\documentclass[twocolumn,traditabstract]{aa}
\usepackage{graphicx,amsmath,amsfonts,amssymb}
\usepackage{epsf}
\usepackage{graphicx}
\usepackage{txfonts}
\usepackage{natbib}
\def\beq{\begin{equation}}
\def\eeq{\end{equation}}

\begin{document}

   \title{High--frequency predictions for number counts and spectral
     properties of extragalactic radio sources. New evidence of a break at
     mm wavelengths in spectra of bright blazar sources.}

   \titlerunning{High--frequency predictions for number counts of ERS}

   \author{Marco Tucci\inst{1} \and Luigi Toffolatti\inst{2,3} \and
     Gianfranco De Zotti\inst{4,5} \and Enrique
     Mart\'\i{nez}--Gonz\'alez\inst{6}}

   \authorrunning{Tucci et al.}

   \institute{LAL, Univ Paris--Sud, CNRS/IN2P3, Orsay, France\\ \and
     Departamento de F\'\i{sica}, Universidad de Oviedo, c. Calvo
     Sotelo s/n, 33007 Oviedo, Spain\\ \and Research Unit associated
     with IFCA-CSIC, Instituto de F\'\i{sica} de Cantabria, avda. los
     Castros, s/n, 39005, Santander, Spain\\ \and INAF--Osservatorio
     Astronomico di Padova, Vicolo dell'Osservatorio 5, 35122 Padova,
     Italy\\ \and International School for Advanced Studies,
     SISSA/ISAS, Astrophysics Sector, Via Bonomea 265, 34136 Trieste,
     Italy\\ \and IFCA-CSIC, Instituto de F\'\i{sica} de Cantabria,
     avda. los Castros, s/n, 39005, Santander, Spain\\}

 \offprints{M. Tucci: tucci@lal.in2p3.fr}

\abstract{We present models to predict high--frequency counts of
  extragalactic radio sources using physically grounded recipes to
  describe the complex spectral behaviour of blazars that dominate
  the mm-wave counts at bright flux densities. We show that simple
  power-law spectra are ruled out by high-frequency ($\nu \ge
  100\,$GHz) data. These data also strongly constrain models featuring
  the spectral breaks predicted by classical physical models for the
  synchrotron emission produced in jets of blazars (Blandford \&
  K\"onigl 1979; K\"onigl 1981). A model dealing with blazars as a
  single population is, at best, only marginally consistent with data
  coming from current surveys at high radio frequencies. Our most
  successful model assumes different distributions of break
  frequencies, $\nu_M$, for BL\,Lacs and flat-spectrum radio quasars
  (FSRQs). The former objects have substantially higher values of
  $\nu_M$, implying that the synchrotron emission comes from more
  compact regions; therefore, a substantial increase of the BL\,Lac
  fraction at high radio frequencies and at bright flux densities is
  predicted. Remarkably, our best model is able to give a very good
  fit to all the observed data on number counts and on distributions
  of spectral indices of extragalactic radio sources at frequencies
  above 5 and up to 220\,GHz. Predictions for the forthcoming sub-mm
  blazar counts from {\it Planck}, at the highest HFI frequencies, and
  from Herschel surveys are also presented.}


   \maketitle
%

\section{Introduction}

In the last few years, large area surveys of extragalactic radio
sources (ERS) at high radio frequencies $\ga10$\,GHz -- the Ryle
telescope 15--GHz 9C survey \citep{wal03}, the Very Small Array (VSA)
one \citep{wat03,cle05} and, especially, the Australia Telescope AT20G
surveys of the southern hemisphere \citep{sad06,mas08,mur10} and the
seven--year all-sky WMAP surveys \citep{gol10} -- have provided a
great amount of new data on number counts, redshift distributions and
emission spectra of radio sources in the $10-100$\,GHz frequency
range, which was poorly explored before.  Moreover, number counts and
related statistics of ERS are now available also at frequencies above
$100$\,GHz, although they are estimated from smaller sky areas and
are consequently limited to fluxes $S\la1$\,Jy \citep{vie09,mar10}.

At the beginning of the year 2011, available data on ERS have
experienced an additional boost, by the publication of the {\it
  Planck's} Early Release Compact Source Catalogue ({\it Planck}
ERCSC) by the {\it Planck} Collaboration \citep{Planck11a}. In fact,
the {\it Planck} ERCSC \citep{Planck11c} comprises highly reliable
samples of hundreds of ERS at bright flux densities and high Galactic
latitude, detected in many of the nine frequency channels
(30--857\,GHz) of the {\it Planck} satellite \citep{tau10}. These
samples of bright ERS -- and the future ones that will be provided by
the {\it Planck} Legacy Catalogue -- will be unrivalled for many years
to come at mm wavelengths. Indeed the statistically complete samples
of ERS selected from the {\it Planck} ERCSC have already allowed the
community to achieve very relevant outcomes on number counts and on
spectral properties in the 30--217 GHz frequency range
\citep{Planck11i}.

Early evolutionary models of radio sources
\citep[e.g.,][]{dan87,dun90,tof98,jac99} were able to give remarkable
successful fits to the majority of data coming from surveys at
$\nu\la10$\,GHz, and down to flux densities of a few mJy.  In
particular, the model by \citet{tof98} was capable to give a good fit
of ERS number counts of WMAP sources as well, albeit with an offset of
a factor of about $0.7$ (see, e.g., \citealt{ben03}). Thus, this model
was extensively exploited to estimate the radio source contamination
of cosmic microwave background (CMB) maps \citep{vie01,vie03,tuc04} at
cm/mm wavelengths. On the other hand, its very simple assumptions on
the extrapolation of source spectra at mm wavelengths as well as new
data published in the last ten years make it currently not
up--of--date for more predictions, although it is still very useful
for comparisons -- even at $\nu\ge100$\,GHz -- after a simple
rescaling \citep[see, e.g.,][]{mar10}.

More recently, the \citet{dez05} and \citet{mas10} new cosmological
evolution models of radio sources have been able to give successful
fits to the wealth of new available data on luminosity functions,
multi-frequency source counts and also redshift distributions at
frequencies $\ga5$\,GHz and $\la5\,$GHz, respectively. These two
models are based on an accurate determination of the
epoch--dependent luminosity functions (l.f.) for different source
populations, usually separated by the value of the spectral index,
$\alpha$, of the observed emission spectrum at low radio frequencies
($1\la\nu\la5\,$GHz) by adopting a simple power-law approximation,
i.e., $S(\nu)\propto \nu^{\alpha}$. This approximation can be well
assumed if there is synchrotron emission, but generally only
in limited frequency intervals.

If $\alpha\ge-0.5$, sources are classified to have a "flat"--spectrum and
are usually divided into flat--spectrum radio quasars (FSRQ) and BL\,Lac
objects, collectively called blazars\footnote{Blazar sources are
  jet-dominated extragalactic objects -- observed within a small angle
  of the jet axis -- in which the beamed component dominates the
  observed emission \citep{ang80}. They are characterized by a highly
  variable and polarized non-thermal synchrotron emission at GHz
  frequencies that is originated from electrons accelerated up to
  relativistic energies in the collimated jets, which come out of the
  central active galactic nucleus \citep[AGN;][]{urr95}.}. Otherwise
they are classified as "steep"--spectrum sources, which are mostly
associated with powerful elliptical and S0 galaxies
\citep{tof87}. Both populations consist of AGN--powered radio
sources and the separation into two main source classes reflects
that the region where the observed radio flux density is
predominantly emitted is different in the two cases. For
steep--spectrum sources, the flux originates in the extended
(optically thin) radio lobes.  For flat-spectrum sources the flux
mainly comes from the compact (optically thick) regions of the radio
jet.  At lower flux densities, different populations of radio
sources also contribute to the population of ``steep''--spectrum
radio sources: dwarf elliptical galaxies, starburst galaxies, faint
spiral and irregular galaxies, etc. (see \citealt{dez05}, for a
thorough discussion on the subject).

The very recent, comprehensive review by \citet{dez10} provides an
up--to--date overview of all data published so far and at the
same time of the cosmological evolution models and of the relevant
emission processes that cause the observed ERS spectra.
Moreover, an interesting and useful discussion of the relevant open
questions on the subject is also presented.

The predictions on high--frequency number counts of ERS provided by
the above quoted cosmological evolution models are based on a
statistical extrapolation of flux densities from the low--frequency
data ($< 5$ GHz) at which the l.f.s are estimated. They adopt a
simple power--law, characterized by an ''average'', fixed, spectral
index, or by two spectral indices (at most) for each source
population. Thus, this ``classical'' modelling has to be considered
as a first -- although successful -- approximation, and it can give
rise to an increasing mismatch with observed high--frequency
($>30$\,GHz) number counts. Indeed, the radio spectra in AGN cores
can be quite different from a single power--law in large frequency
intervals. In particular, a clear steepening (or a bending down) at
mm wavelengths is theoretically expected \citep{kel66,bla79}. This
steepening has been already observed in some well known blazars
\citep{cle83} and has also been statistically suggested by recent
analyses of different ERS samples at $\nu> 30$ GHz \citep[see,
  e.g.,][]{wal07,gon08}.

The scenario giving rise to a moderate or to a more relevant
($\Delta\alpha>$ 0.5) spectral steepening in blazars -- at frequencies
of tens to hundreds GHz -- is complex because different physical
processes are intervening at the same time. The more relevant ones
for our purposes are briefly sketched below (and will be discussed
more extensively in Section 4).

{\bf a)} In the inner part of the AGN jet, i.e. typically at several
thousand Schwarzschild radii from the central collapsed object
\citep{ghi09}, various different and self-absorbed source components
("spherical blobs", very probably
accelerated by shock waves in the plasma; \citealt{bla79}) are
responsible of the observed synchrotron radiation at different
distances from the central AGN core \citep{mar85,mar96}
\footnote{Various examples of blazar spectra tentatively fitted by
  different synchrotron components in the AGN jet are also given by
  \citet{Planck11k}}. In this situation, the emerging jet-- and blazar
spectrum is approximately flat ($\alpha\approx 0.0$, but it can be
either very moderately steep, or even inverted); this is an indication
that {\it ``the cores are partially optically thick to synchrotron
  self-absorption, as expected in the jet model''} \citep{mar96}.

{\bf b)} However, the high--energy (relativistic) electrons --
responsible for the self-absorbed synchrotron radiation, with the
quoted almost flat emission spectrum \citep{kel69,mar80a} --
injected into the AGN jets are loosing their energy because of
synchrotron losses, thus cooling down (i.e., electron ageing) when
the rate of injection is not sufficient to balance the radiation
losses. As a consequence, their synchrotron radiation spectrum has
to steepen (at a given frequency) to a new $\alpha$ value, which
depends on whether the electrons are continuously or instantaneously
injected \citep[see, e.g.,][Section II and Figure 5]{kel66}.

{\bf c)} In standard models of the synchrotron emission in blazar
jets, the size of the optically thick core of the jet varies with
frequency as $r_c\propto \nu_s^{-1/k}$ \citep[see,
  e.g.,][]{kon81,cle83,lob10,sok11}, corresponding to the smallest
radius, $r_M=r_c$, from which the optically thin synchrotron emission
from the jet can be observed (see Sect. 4).\footnote{Recently, Very
    Long Baseline Array (VLBA) simultaneous observations of the
    frequency--dependent shift seen in the core position of blazar
    sources (known as the "core shift") at nine frequencies in the
    1.4--15.4 GHz range \citep{sok11} have found new evidence of a
    parameter value $k\approx 1$ in the above quoted law, confirming
    theoretical predictions.} Therefore, the observed blazar spectrum
  at $\nu\geq\nu_M$, now in the optically thin regime \citep[see,
    e.g.,][]{cle83}, will be moderately steep, with a typical
  $\alpha\geq\alpha_0+0.5$ value determined by synchrotron-radiation
  losses owing to electron ageing \citep{kel66}.

As a result, at frequencies in the range $10-1000$ GHz, depending on
the parameters more relevant for the physical processes
discussed here (see Sect. 4), the spectra of blazar sources have to show
a {\it break} or a {\it turnover} \citep[see, e.g.][]{kon81,mar96}
with a clear spectral steepening at higher frequencies.

Moreover, in the frequency range where the CMB reaches its maximum,
the interest in very accurate predictions of number counts of radio
ERS is obviously increased, given that they constitute the most
relevant contamination of CMB anisotropy maps on small angular
scales \citep[see, e.g.,][]{tof99,tof05}. Indeed, current
high--resolution experiments at mm/submm wavelengths require not
only to remove bright ERS from CMB anisotropy maps but also to
precisely estimate the contribution of undetected point sources to
CMB anisotropies. This problem is particularly relevant for current
as well as forthcoming CMB polarization measurements
\citep{tuc04,tuc05}.

In the present work we still compare our predictions on number
counts and other statistics of ERS to the most relevant
observational data sets at $\nu>5$\,GHz by extrapolating the
spectral properties of the ERS observed at low radio frequencies.
However, our approach is somewhat different than before. The
fundamental difference consists in providing, for the first time,
a statistical characterization of the sources' spectral behavior
at high radio frequencies that takes into account -- at least for the
most relevant population of ERS at cm to mm wavelengths -- the main
physical mechanisms responsible of the emission. We focus, in
particular, on flat--spectrum radio sources, given that they are the
dominant source population in the flux-- and frequency range we are
interested in \citep{gio04}.

Finally, we analyse current data of another source population,
although less relevant at bright flux densities: the population of
the so-called "inverted"--spectrum sources (i.e., with a positive
value of the spectral index $\alpha$) detected in high--frequency
radio surveys \citep{dal00,bol04,sad06,mur10}. We give
predictions also on their contribution to number counts and to other
related statistics. In this latter case, our approach is still a
purely statistical one, without dealing with the physical conditions
that cause the observed emission. This simplified choice is
justified by the current lack of data on the underlying physics of
this source population and because these sources can be
substantially contaminated by blazars observed in their active phase
\citep[see, e.g.,][]{Planck11k,Planck11j}.

The outline of the paper is as follows: in Sect. 2 we discuss number
counts of flat-- and steep--spectrum radio sources at 5 GHz; in Sect.
3 we analyse the spectral index distribution of the different ERS
populations; in Sect. 4 the basic assumptions of the simplified
physical model for flat--spectrum sources is presented and extensively
discussed; in Sect. 5 we separately present the data on
high--frequency peak spectrum (GPS) sources; in Sect. 6 we summarize
the different model assumptions for the various source populations we
identify in Sects. 3, 4, and 5; in Sect. 7 we present and discuss our
model predictions on the extrapolation of the 5 GHz number counts and
spectral properties of ERS to higher radio frequencies; finally, in
Sect. 8 we summarize our main conclusions. In addition, and for
reducing the main body of the article, we refer the reader to three
appendices: in Appendix A we give a brief description of the data sets
currently available on ERS and analysed in this paper; the simplified
formula adopted here for the estimation of the break frequency in
blazars spectra is presented in Appendix B; finally, we discuss the
main physical quantities that determine the value of the break
frequency in Appendix C.


\section{Number counts of extragalactic radio sources at 5\,GHz}
\label{s1}

\begin{figure}
\centering
\includegraphics[width=9cm]{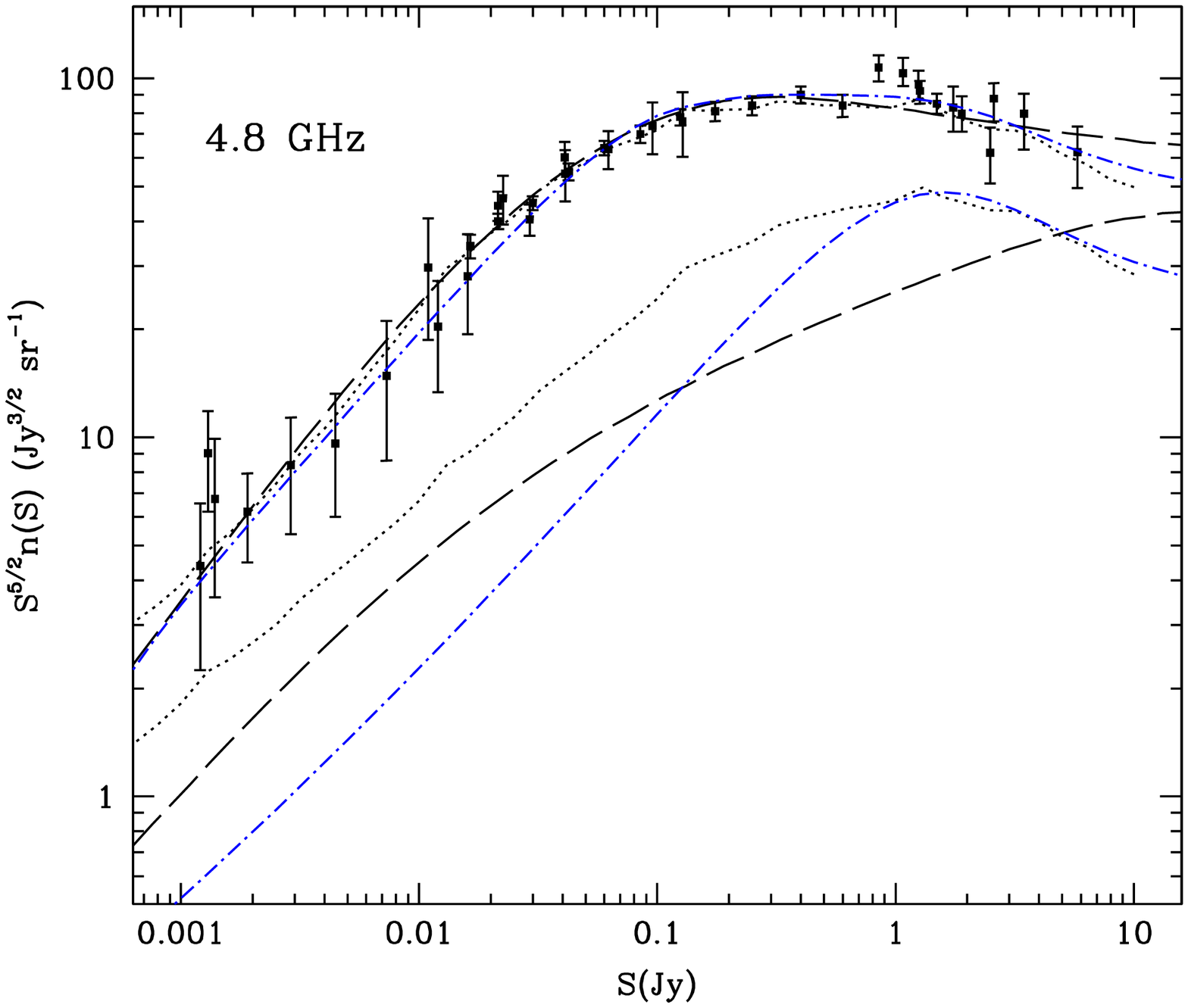}
\caption{Differential number counts normalized to $S^{5/2}$ for ERS at
  4.8\,GHz. Filled squares are observational data (see De Zotti et
  al. 2010). The plotted curves represent the estimated number counts
  from the models by \citet[][dotted line]{tof98}, \citet[][blue
    dash-dotted line]{dez05}, and \citet[][dashed line]{mas10}. Lower
  curves with the same line types as above represent the estimated
  number counts of flat--spectrum sources given by the same models.}
\label{f1}
\end{figure}

The differential number counts of ERS at $\sim5$\,GHz are well known
and have been extensively analysed and discussed
\citep[see][]{dez10}. We plot in Fig.\,\ref{f1} the observed number
counts with the fits yielded by different models, i.e.,
\citet{tof98}, \citet{dez05}, and \citet{mas10}. The plot covers the
flux range between about 1\,mJy and 10\,Jy. At these levels the
number counts are essentially dominated by AGNs, while at fainter
fluxes the contribution of star--forming galaxies becomes
increasingly important \citep[see, e.g.,][and references therein for
a thorough discussion on
  the subject]{mas10}.

However, the knowledge of the total counts of ERS at GHz frequencies
is not sufficient for making predictions at higher radio
frequencies. Because of the presence of different source populations
with different spectra, it is necessary to identify which populations
dominate the number counts and their relative number as a function of
the flux density. To this aim, the large--area surveys of ERS
available at GHz frequencies help us in classifying the observed ERS
as steep-- or flat--spectrum sources\footnote{Hereafter, we adopt the
  usual convention $S(\nu)\propto\nu^{\alpha}$. Thus, we classify an
  ERS as steep--spectrum if $\alpha<-0.5$, and as flat--spectrum if
  $\alpha\ge-0.5$, by using the flux densities measured at 1.4 and
  4.8 GHz. Moreover, we consider as inverted--spectrum all those ERS for
  which $\alpha\ge0.3$.}.

\begin{figure*}
\centering
\includegraphics[width=85mm]{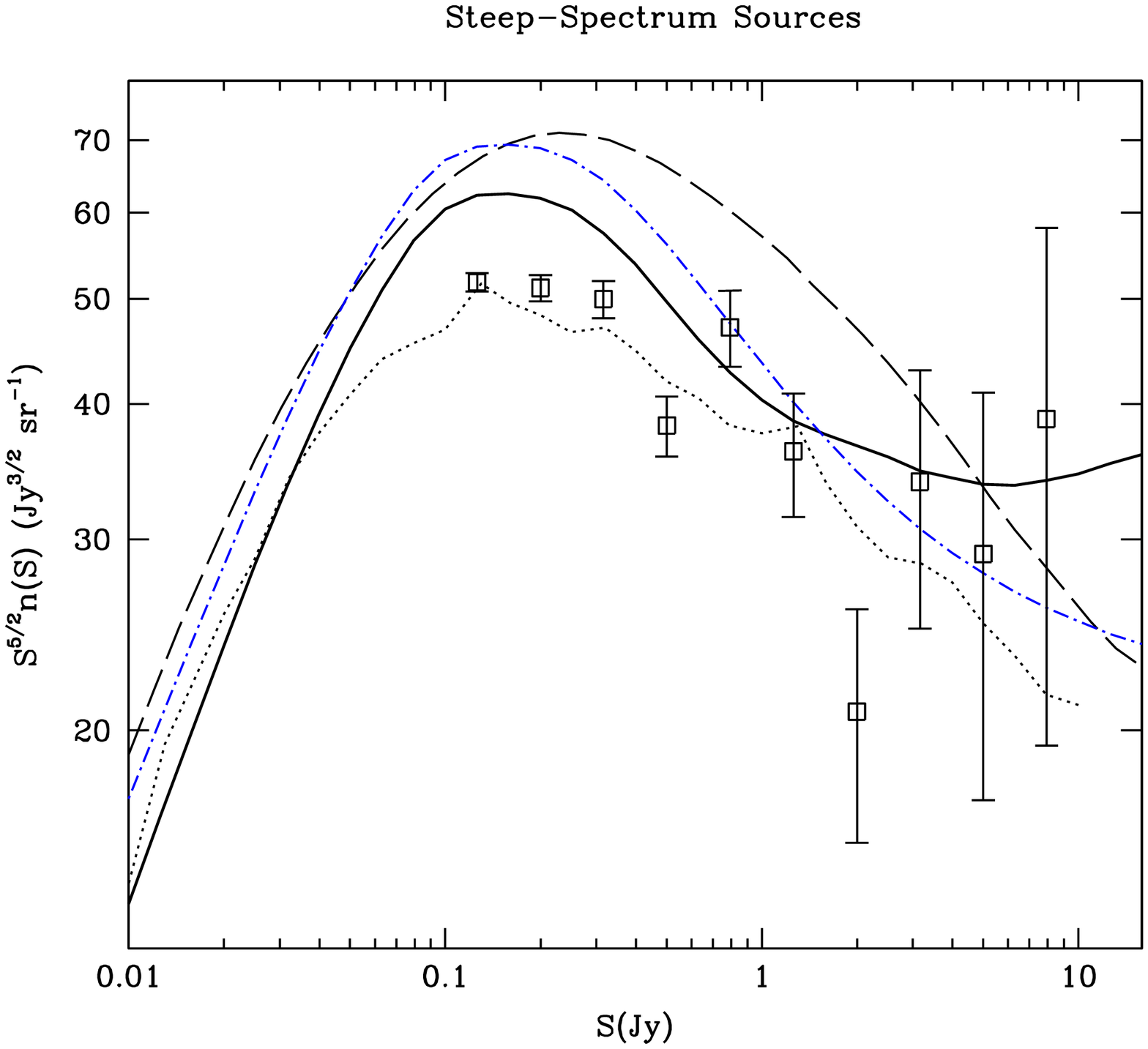}
\includegraphics[width=85mm]{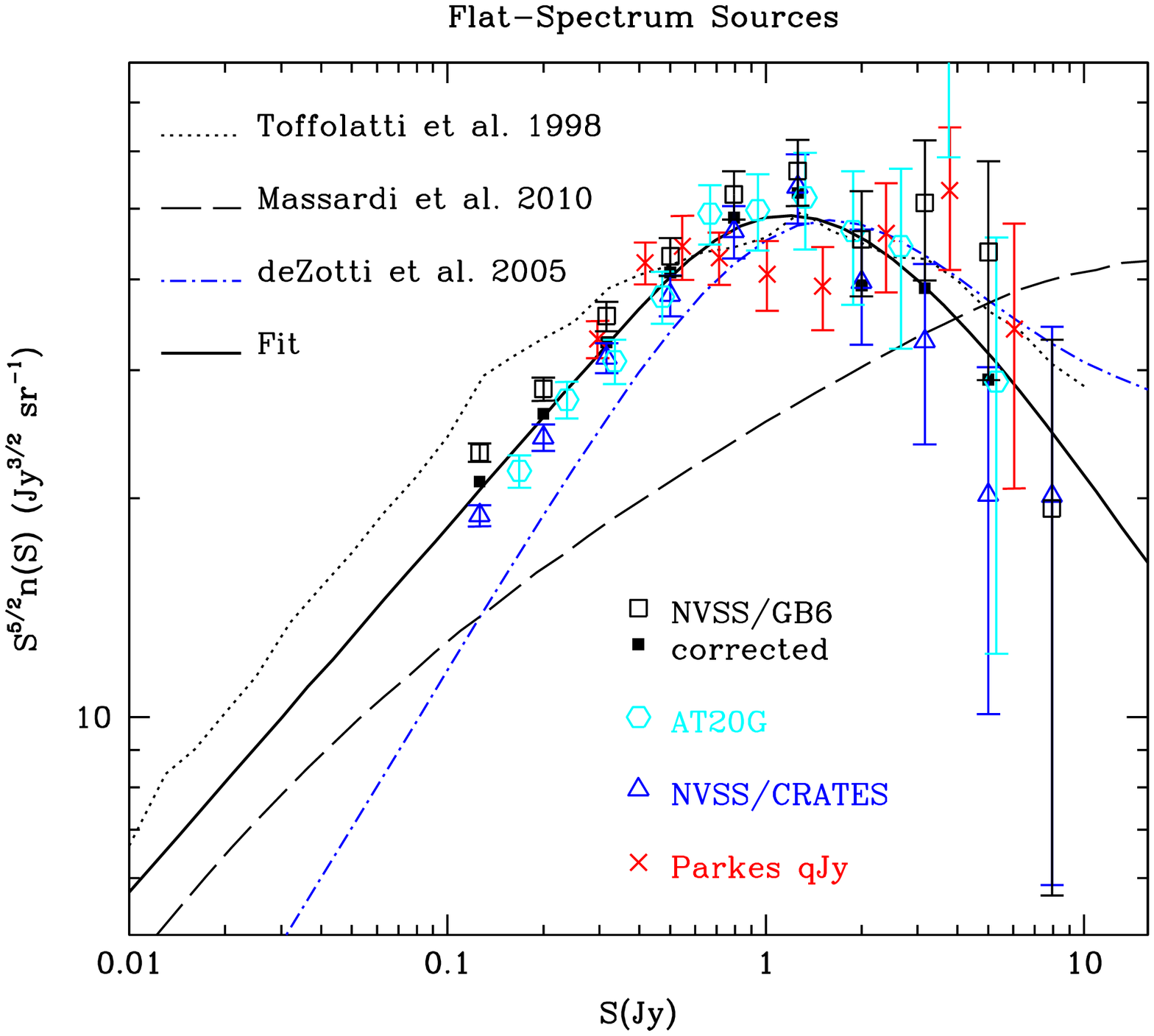}
\caption{Differential number counts at 4.8\,GHz of steep--spectrum
  ({\it left panel}) and flat--spectrum ({\it right panel}) ERS from
  the NVSS/GB6 data (black empty squares). The predictions from the
  cosmological evolution models plotted in Fig.\,\ref{f1} are
  represented here by the same line-types as in that figure. In the
  {\it left panel} the thick continuous line represents the {\it
    difference} between the total number counts of the \citet{tof98}
  evolution model and our best fit to the observed number counts of
  flat--spectrum ERS (right panel; thick continuous line), i.e., it
  represents our best estimate of the differential number counts of
  steep--spectrum ERS. In the {\it right panel} we also plot the
  number counts of flat--spectrum ERS estimated from i) NVSS/GB6 data
  but corrected for resolution effects (black solid squares); ii) the
  \citet{jac02} sample (red astericks); iii) the CRATES sub--sample in
  the GB6 sky area (blue empty triangles); iv) the 5--GHz sample of
  AT20G sources (cyan empty circles). See Sec.\ref{s1}\ for more
  details.}
\label{f2}
\end{figure*}

In Fig.\,\ref{f2} we plot the differential number counts of these
two source populations as calculated from different source samples,
which are described in detail in Appendix\,\ref{s0}. These results require a
careful discussion because the simple estimates of source spectral
indices are not free from uncertainties, even in the case of
negligible errors on published flux densities. First of all, we
considered the catalogue of ERS with spectral information based on
the NVSS and GB6 surveys (hereafter, the {\bf NVSS/GB6}
sample), which is defined in Appendix\,\ref{s0}. Spectral indices from
NVSS and GB6 data should be taken with some caution because of the
different measurement epoch and the different resolution of the
antennae. In particular, biased values of spectral indices could
arise for resolution effects, especially when GB6 objects are
resolved by the NVSS or have multiple components in the NVSS, which
causes a flatter spectral index. 
On the other hand, variability mostly affects flat--spectrum
sources. We discuss this specific problem in
Section\,\ref{ss1}.

For estimating the magnitude of resolution effects on the
4.8--GHz number counts of flat--spectrum sources, we identified all
ERS resolved by the NVSS (i.e., with major axis
$>45$\,arcsec) that are classified as flat--spectrum sources. We found 173
resolved objects. Because flat--spectrum sources are usually
associated with compact objects with an emission dominated by the AGN
core, these sources could be false identifications, and we
marked them as steep--spectrum sources. Then we
dealt with flat--spectrum sources with multiple NVSS counterparts and
without clearly dominant NVSS objects: we looked for objects the
brightest NVSS counterpart of which contributes $<75$\% of the total
flux density and found 102. We redistributed these
sources between the two populations according to the proportion of
flat-- and steep--spectrum sources in flux--density bins. After
these corrections, the number counts of flat--spectrum sources were
significantly reduced, especially at fluxes $S\la300\,$mJy, whereas
at higher fluxes the correction was small (see
Fig.\,\ref{f2}).\footnote{These corrected counts will be not used
later. However, they are an useful indication of the possible
uncertainty in the spectral classification because of the different
resolution between the two surveys discussed here.}

Number counts of flat--spectrum sources in Fig.\,\ref{f2} were also
obtained from the Parkes quarter--Jy sample \citep{jac02}. 
The Parkes beam size is 8\,arcmin at 2.7\,GHz and 4\,arcmin at
5\,GHz, i.e., a factor 2 larger at the lower frequency. Therefore,
resolution effects should go in the opposite direction with respect
to the NVSS/GB6 sample. 
Moreover, we increased the number counts of these sources by
10\% (the percentage observed in the NVSS/GB6 sample) to
take into account sources with $-0.5\le\alpha\le-0.4$, which were 
excluded from the original sample.

Fig.\,\ref{f2} also shows the results from the CRATES catalogue
\citep{hea07} in the area of the GB6 survey. There, the number counts
is computed taking into account 8.4--GHz measurements: we calculated
the spectral index between 1.4 and 8.4\,GHz, and then the differential
number counts at 4.8\,GHz only for those sources that verified the
condition $\alpha_{1.4}^{8.4}\ge-0.5$. The new source counts must be
considered a lower limit because some really flat--spectrum sources
could have been discarded owing to variability or a spectral
steepening between 4.8 and 8.4\,GHz. We do not expect many of these,
whereas the number of false identifications coming from the
classification of NVSS/GB6 data should be strongly reduced. If we
compare these number counts with the ones derived from the NVSS/GB6
sample, we see that they partially agree at flux densities
$S\ga0.5$\,Jy, whereas the discrepancy increases with lower fluxes.

Finally, we also plotted the number counts of flat--spectrum
sources estimated from the 5--GHz measurements present in the
Australia Telescope Compact Array 20\,GHz survey
\citep[AT20G;][]{mur10}: we limited our analysis to the
almost--complete samples at declination $\delta<-15^{\circ}$ and
flux limits 100\,mJy and 50\,mJy, respectively (indicated as {\bf
AT20G--d15S100} and {\bf
  AT20G--d15S50}, see Appendix\,\ref{s0}); then, we took all
sources with i) measurements at 5 and 8\,GHz; ii) 5--GHz flux density
$S_5\ge100$\,mJy; iii) $\alpha_5^8\ge-0.5$. The resulting number
counts were corrected for the completeness of the samples and for the
percentage of sources with 5-- and 8--GHz data (89\% and 84\%
respectively in the two AT20G sub--samples). We observe that the
number counts agree well with previous counts at
$S\ga0.5$\,Jy, and with CRATES data at lower fluxes. ATCA measurements
have the advantage to be nearly simultaneous and made with antennae of
similar resolutions, which reduces therefore the uncertainty in the
spectral index estimates. Although not complete at 5\,GHz, the sample
built in this way provides reliable results starting from
$S\ga200$\,mJy, which should be considered at least as a lower limit
for fluxes $S<1\,$Jy.

In Fig.\,\ref{f1}--\ref{f2} the differential number counts predicted
from the cosmological evolution models of \citet{tof98},
\citet{dez05}, and \citet{mas10} are also displayed. We see that
whereas these models fit the total number counts at 5\,GHz extremely
well, their predictions on the number of flat-- or steep--spectrum ERS
show significant differences and in general disagree with the
published survey data. The \citet{tof98} and \citet{dez05} models fit
observational counts of the two populations at flux $S>0.5$\,Jy quite
well, but at lower fluxes they tend to overestimate (in the former
case) or to underestimate (in the latter case) the number of
flat--spectrum sources. On the other hand, the ratio of steep-- and
flat--spectrum sources expected from the \citet{mas10} model seems to
be quite far from the observed ones, at least for $S\ge0.1$\,Jy. It is
unlikely that uncertainties in the classification of radio sources
could explain this great discrepancy.

As a conclusion, at flux densities $S\ge100$\,mJy we can fit the
differential number counts of flat--spectrum sources obtained from
the observational data by a broken power law \beq
\bar{n}(S)=n_0{(S/S_0)^k \over
  1-e^{-1}}\Big(1-e^{-(S/S_0)^{l-k}}\Big)\,,
\label{e1}
\eeq
where $\bar{n}(S)$ is the differential number counts normalized to
$S^{5/2}$. We find $n_0=47.4$\,Jy$^{-1}$sr$^{-1}$, $S_0=1.67$\,Jy,
$k=0.50$ and $l=-0.66$. We extrapolate this curve also to
$S<100$\,mJy (represented as a thick continuous line in the right
panel of Fig.\,\ref{f2}).

\begin{table*}
\caption{Spectral properties of sources in the NVSS/GB6 sample. For
  each flux density range the columns indicate the total number of
  sources, the number of steep-- and flat--spectrum (+inverted)
  sources, and the median values of their corresponding spectral index
  distributions; in the last column, the number of inverted--spectrum
  sources ($\alpha\ge0.3$) is also given.  $\alpha_{\rm peak}$ and
  $\sigma_{\alpha}$ indicate the parameters of the best--fit truncated
  Gaussians for steep-- and flat--spectrum sources (see Section 3.2).}
\centering
\begin{tabular}{lccccccccccc}
\hline
Flux density & & \multicolumn{4}{l}{Steep} & &
\multicolumn{4}{l}{Flat(+Inverted)} & Inverted \\
(mJy) & N$_{tot}$ & N & median & $\alpha_{\rm peak}$ & $\sigma_{\alpha}$ & & N &
median & $\alpha_{\rm peak}$ & $\sigma_{\alpha}$ & $N$ \\
\hline
$[$100,\,158) & 3832 & 2662 & -0.87 & -0.87 & 0.15 & & 1170 & -0.14 & -0.39 &
  0.38 & 167 \\
$[$158,\,251) & 2068 & 1335 & -0.87 & -0.86 & 0.14 & & 733  & -0.11 & -0.29 &
    0.37 & 103 \\
$[$251,\,400) & 1125 & 657  & -0.85 & -0.86 & 0.14 & & 468  & -0.09 & -0.17 &
    0.30 & 61 \\
$\ge400$      & 1102 & 498  & -0.81 & -0.81 & 0.14 & & 604  & -0.03 & -0.14 &
    0.40 & 117 \\
\hline
\end{tabular}
\label{t1}
\end{table*}


\section{Spectral indices of radio sources}
\label{s2}

\subsection{Spectral index distributions in the NVSS/GB6 sample}
\label{ss1}

It is usually assumed that the spectral index distribution of ERS at
GHz frequencies can be described in all flux density ranges by the
sum of two Gaussian distributions with maximum at $\alpha\sim-0.8$
(for steep--spectrum sources) and at $\alpha\sim0.0$ (for
flat--spectrum sources; see, e.g., \citealt{dez10}).
For checking this assumption, we calculated the spectral index
distributions at 5\,GHz of steep-- and flat--sources in the NVSS/GB6
sample by dividing them into four flux--density intervals (see
Table\,\ref{t1} and Fig.\,\ref{f4}). On one hand, the distributions
calculated for steep--spectrum sources seem not to change
significantly with the flux density, peaking around $-0.9$ and
having a median spectral index between $-0.81$ and $-0.87$. On the
other hand, the distributions calculated for flat--spectrum sources
show a clear dependence on the flux density interval, with steeper
spectra at lower flux densities. For $S>400$\,mJy the maximum in the
distribution is around $\alpha\sim 0$, which agrees well with
observations from other surveys (e.g., \citealt{ric04}). However, at
lower flux--density ranges, the maximum in the distributions
gradually shifts to lower values of $\alpha$ and is around
$-0.4$ for $S<251\,$mJy. The relative number of inverted--spectrum
sources ($\alpha\ge0.3$) is about 14\% at low flux densities, but it
increases to almost 20\% at $S>400$\,mJy.

\subsection{Spectral index distributions vs. variability}
\label{ss1a}

Because spectral indices are computed from non--simultaneous
observations at 1.4 and 4.8\,GHz, the source's variability could affect
the spectral index distributions, especially for flat--spectrum sources
that are observed to be variable at low frequencies as well. Accordingly, we
expect that source variability will induce a higher dispersion in
spectral index distributions and, related to it, false identifications
between steep-- and flat--spectrum source populations.

In \citet{tin03} a study of source variability at GHz frequencies is
carried out for 185 sources (167 flat-- and 18 steep--spectrum ERS):
observations were made over a 3.5 year period with the ATCA at 1.4,
2.5, 4.8 and 8.6\,GHz. For each source the authors provide the variability
index, defined as the RMS fractional variation from the mean flux
density. They found that at 4.8\,GHz $\sim40$\% of flat--spectrum
sources show a variability $>10$\%, whereas only 5\% vary by
$>30$\%. Variability at 1.4\,GHz is slightly lower. On the other
hand, no significant variability is observed in steep--spectrum
sources, except for five sources (over 18) with a variability index
between 4 and 20\,per\,cent. These results are consistent with
variability studies carried out at higher frequencies
\citep{sad06,bol06}.

\begin{figure}
\centering
\includegraphics[width=9cm]{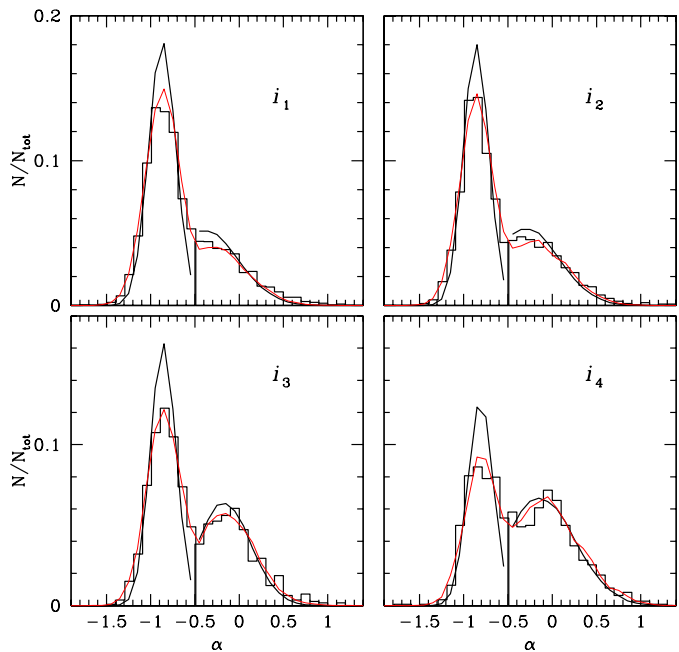}
\caption{Distributions of the spectral index calculated from the
  NVSS/GB6 sample for different flux density intervals: $i_1$)
  $S=$[100,\,158]\,mJy; $i_2$) $S=$[158,\,251]\,mJy; $i_3$)
      $S=$[251,\,400]\,mJy; $i_4$) $S\ge$400\,mJy. The thick black
        lines are the ``best--fit'' truncated Gaussians, and the thin
        red lines are the total distributions obtained from them after
        introducing source variability (see the text).}
\label{f4}
\end{figure}

We assume that flux densities vary at 1.4 and 5\,GHz according to
the Tingay et al.  results, which we summarize in Table\,\ref{t1b}.
We generated a simulated sample of steep-- and flat--spectrum
sources at 5\,GHz with a flux density distribution consistent with
the GB6 number counts and with spectral index distributions given
by Gaussian functions truncated at $\alpha=-0.5$ (with peak position
$\alpha_{\rm peak}$ and dispersion $\sigma_{\alpha}$). The 1.4--GHz
flux density is calculated as $S_5(1.4/5)^{\alpha}$. After
introducing variability in flux densities at 1.4 and 5\,GHz, we
computed the new $\alpha$--distributions by taking into account
the uncertainties in GB6 and NVSS flux densities as well. Finally, we found
the parameters of the truncated Gaussian distributions ($\alpha_{\rm
peak}$ and $\sigma_{\alpha}$), for which the simulated
$\alpha$--distributions give the best fits of the observed NVSS/GB6
distributions. The results are reported in Table\,\ref{t1} and in
Fig.\,\ref{f4}, where we show both the original truncated Gaussians
and the final $\alpha$--distributions produced from them. The value
of $\alpha_{\rm peak}$ is $-0.14$ for $S\ge400$\,mJy, and steadily
decreases to $-0.4$ in the lowest flux--density range. We see that
simulations can reproduce the NVSS/GB6 results very well.
Variability only partially affects the spectral index distributions,
increasing their dispersion, and makes some flat--spectrum sources
move to the steep--spectrum class. This effect is particularly
relevant for sources with spectral index close to $-0.5$ and at flux
densities $<250\,$mJy.

The previous simulations also give us the number of sources that are
misclassified because of variability. We found that the number of
flat--spectrum sources from the NVSS/GB6 sample can be
underestimated by about 5--8\,per\,cent, whereas the number of
inverted--spectrum sources can be overestimated by only a few
per\,cent. Owing to their very low variability, the number of
steep-spectrum ERS classified as flat--spectrum sources is small,
and only down to $S\sim100$\,mJy can partially compensate errors
in the classification of flat-spectrum ERS.

\begin{table}
\caption{Variability indices at 1.4 and 4.8\,GHz for flat-- and
  steep--spectrum sources. Columns 2, 3, and 5 give the fractions of
  sources that show the corresponding variability index in columns 1
  and 4. For steep-spectrum sources there is no significant difference
  between the variability indices at 1.4 and 4.8 GHz. The estimated
  values are from \citet[][see text]{tin03}.}  
\centering
\begin{tabular}{ccccccc}
\hline
Variab. & & \multicolumn{2}{l}{Flat} & & Variab. & Steep \\
index & & 1.4\,GHz & 4.8\,GHz & & index & 1.4/4.8\,GHz \\
\hline
$<10\%$ & & 0.69 & 0.60 & & $<4\%$ & 0.70 \\
10--20\% & & 0.22 & 0.23 & & 4--20\% & 0.30 \\
20--30\% & & 0.07 & 0.12 & & & \\
30--50\% & & 0.02 & 0.05 & & & \\
\hline
\end{tabular}
\label{t1b}
\end{table}

\subsection{High--frequency steepening of steep--spectrum sources}
\label{ss2a}

A high--frequency spectral steepening is expected in steep--spectrum
sources due to electron ageing \citep{kel66} and has been observed in
multifrequency surveys at $\nu>5$\,GHz
\citep{bol04,ric06}. \citet{ric06} compared the 2.7--5\,GHz and
5--18.5\,GHz spectral indices and found a median steepening of
$\Delta\alpha=0.32$.

\begin{figure}
\centering
\includegraphics[width=8.5cm]{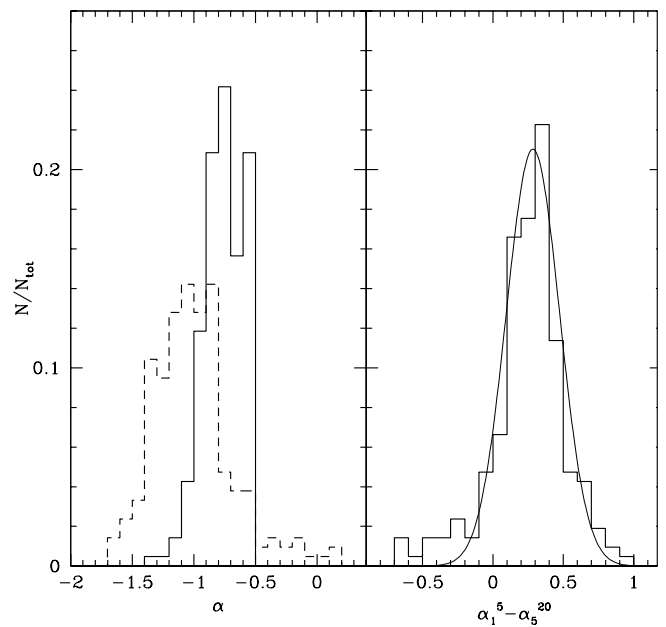}
\caption{{\it Left panel}: Distributions of 1--5\,GHz (solid line) and
  5--20\,GHz (dashed line) spectral indices of steep--spectrum sources
  selected from the AT20G catalogue. {\it Right panel}: distribution
  of the difference $\alpha_1^5-\alpha_5^{20}$. The thin line is the
  Gaussian fit.}
\label{f5}
\end{figure}

A spectral steepening in this frequency range is also observed in
the AT20G data: we selected sources in AT20G--d15S50 with 5--GHz
flux density $S\ge500$\,mJy and spectral index $\alpha_1^5<-0.5$. We
found 211 objects. In Fig.\,\ref{f5} we plot the spectral index
distribution computed between 1--5\,GHz and 5--20\,GHz, and the
difference $\Delta\alpha=\alpha_1^5-\alpha_5^{20}$. A median
steepening of 0.28 is found, in agreement with the \citet{ric06}
result. The distribution of $\Delta\alpha$ is well described by a
Gaussian with $<\Delta\alpha>=0.28$ and $\sigma\simeq0.20$, except
for the negative tail. Values of $\Delta\alpha<0$ (see also the
presence of sources with $\alpha_5^{20}>-0.5$ in the left panel of
Fig.\,\ref{f5}) can arise both from sources with spectra that
flatten \citep{tuc08} and also from source variability because
1\,GHz data are not simultaneous with the AT20G measurements.

At frequencies $\nu>20\,$GHz a further spectral steepening is still
possible: a recent indication in this sense is coming from the
follow--up of 9C sources \citep[see][]{wal07}.

\subsection{High--frequency spectral behaviour of flat--spectrum
  sources}
\label{ss2}

\begin{figure}
\centering
\includegraphics[width=8.5cm]{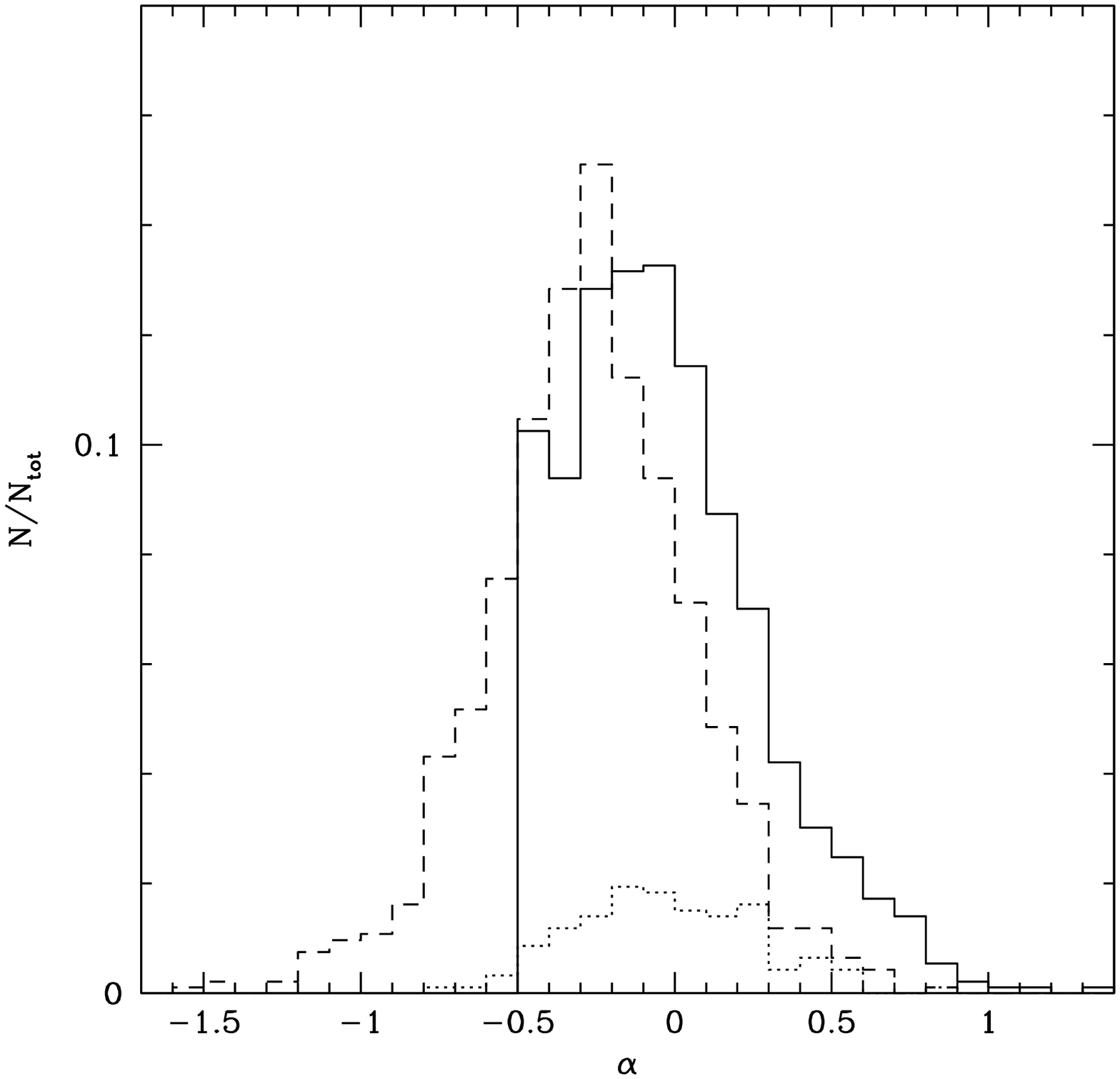}
\includegraphics[width=8.5cm]{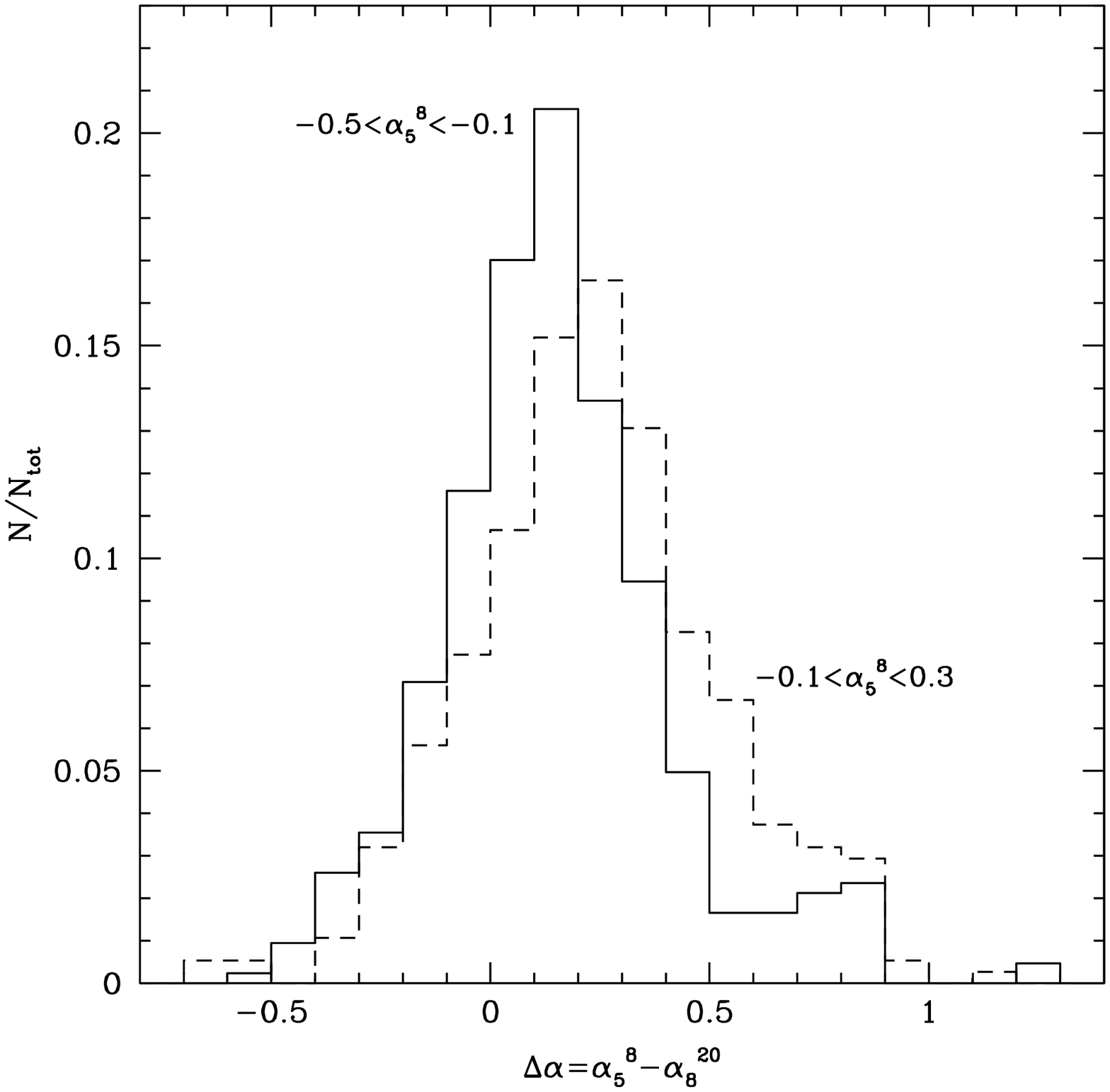}
\caption{{\it Upper panel:} distributions of spectral indices
  $\alpha_5^8$ (solid line) and $\alpha_8^{20}$ (dashed line) for
  sources selected in AT20G--d15S50 with $S_5\ge200$\,mJy and
  $\alpha_5^8\ge-0.5$. The dotted line shows the distribution of
  $\alpha_8^{20}$ for inverted--spectrum sources
  ($\alpha_5^8\ge0.3$). {\it Lower panel:} distribution of
  $\Delta\alpha$ for sources with $-0.5\le\alpha_5^8<-0.1$ (solid
  line) and with $-0.1\le\alpha_5^8<0.3$ (dashed line).}
\label{f4a}
\end{figure}

We focus now on the spectral behaviour of flat--spectrum sources at
frequencies $\nu>5\,$GHz. The AT20G survey, thanks to its
near--simultaneous observations at 5, 8 and 20\,GHz, allows us to
study in a quite broad range of frequencies spectral shapes that
should be not significantly affected by variability \citep{mas10b}. To
this aim, we selected the sources of the AT20G--d15S50 sample with
$S_5\ge200\,$mJy and with 5--8\,GHz spectral index
$\alpha_5^8\ge-0.5$. This sample of 798 objects is about 90\% complete
for sources with $\alpha_8^{20}\ga-0.5$, but it can lose objects that
undergo spectral steepening between 8 and 20\,GHz. In Fig.~\ref{f4a}
we show the distributions of $\alpha_5^8$ and $\alpha_8^{20}$ for
those objects. The distribution of $\alpha_8^{20}$ is shifted to lower
values by about 0.1/0.2 with respect to the $\alpha_5^8$ distribution,
with a large tail of objects with steep spectral index, extending up
to $-1.5$. The percentage of sources with $\alpha_8^{20}<-0.5$ is
$\sim22\%$. The median of distributions, excluding inverted--spectrum
($\alpha_5^8\ge0.3$) sources, is $-0.08$ and $-0.27$ for 5--8\,GHz and
8--20\,GHz spectral indices, respectively. Fig.~\ref{f4a} also shows
the histogram of $\Delta\alpha=\alpha_5^8-\alpha_8^{20}$. In this case
we separate sources with 5--8\,GHz spectral indices that are
moderately steep ($-0.5\le\alpha_5^8<-0.1$) and flat
($-0.1\le\alpha_5^8<0.3$). We found that sources with flatter
5--8\,GHz spectral index tend to have a more relevant steepening
between 8 and 20\,GHz (the medians $\Delta\alpha$ are 0.13 and 0.23,
respectively). An interesting feature in both distributions is also
the presence of sources with typical steepening around 0.7--0.9, which
seem to produce a secondary peak in the $\Delta\alpha$ distribution of
sources with $-0.5\le\alpha_5^8<-0.1$. Because of the incompleteness
of the sample for objects with $\alpha_8^{20}<-0.5$, the peak could be
more prominent than observed, and it could extend to
$\Delta\alpha\ga1$. Finally, we note that most inverted--spectrum
sources between 5 and 8\,GHz ($\alpha_5^8\ge0.3$) have
$\alpha_8^{20}<0.3$, and only $\sim12\%$ are still inverted up to
20\,GHz.

Our findings agree well with the results of
\citet{mas10b}. A very strong average spectral steepening has also
been found in the bright PACO sample \citep{mas10c}, where the
typical differences between the 5--10\,GHz and 30--40\,GHz spectral
indices are in the range 0.5--1.


Additional indications of a spectral steepening in flat--spectrum
sources at frequencies higher than 20\,GHz are given by OCRA--p
observations at 31\,GHz of CRATES selected sources \citep{pee10} and
by ATCA observations at 95\,GHz of 130 AT20G sources \citep{sad08}.
These results are also supported by WMAP data \citep{gon08} at Jy
levels, and by ACT data at 148\,GHz \citep{mar10}.

\section{A spectral model for flat--spectrum sources}
\label{s3}

As discussed in the previous section, recent multi-frequency surveys
have clearly shown that ERS with flat spectra at GHz frequencies can
present a downwards spectral curvature already at few tens of GHz.
This spectral behaviour could well be an indication that the
transition in the synchrotron spectra from the optically--thick to
the optically--thin regime is observed, and that this transition can
occur even at cm wavelengths. This conclusion is also supported
by the results on the compactness of radio sources as a function of
the spectral index, obtained from the AT20G survey. As an example,
\citet[][see their Fig.\,4]{mas10b} do find a remarkable separation
between steep--spectrum ($\alpha_1^5<-0.5$) ``extended'' sources and
flat--spectrum ($\alpha_1^5\ge-0.5$) ``compact'' sources. However,
if the compactness is considered versus the spectral index between 8
and 20\,GHz, there appears to be also a population of compact and
steep--spectrum ERS, corresponding to the 12\% of the sample, which
is not present at lower frequencies and suggests a spectral
steepening at $\nu<$20\,GHz.

By following this hypothesis, the key point for making predictions
on the high--frequency contribution of flat--spectrum ERS to number
counts is to make an estimate of the frequency at which the spectrum
of this source population can experience a substantial steepening
($\Delta\alpha\ga0.5$). To this aim, we use a simple physical
description of the AGN radiation emission in the inner part of their
relativistic jets. Our predictions are then calculated on the basis
of typical physical parameters in current models for the emission of
blazar sources.

\subsection{Break frequency in synchrotron spectra from relativistic
  blazars jets}
\label{sect:break}

It is generally believed that flat spectra observed in radio sources
result from the superposition of different components of relativistic
jets, each with a different synchrotron self--absorption frequency
\citep{kel69,cot80}. Relativistic shocks in the jet produce flares
that are responsible for radio flux density variations on intervals
from months to years \citep[e.g.,][]{mar85,val92}.

Flat radio spectra are predicted by models of synchrotron emission
from inhomogeneous unresolved relativistic jets \citep{bla79,kon81}.
According to them, the observed flux density at a given frequency is
dominated by the synchrotron--emitting component, which becomes
self--absorbed at that frequency, because synchrotron self-absorption
is the most likely absorption mechanism acting in the compact conical
jet \citep{kon81}. When moving to high frequencies, self--absorbed
synchrotron emissions from more and more compact regions are thus
observed closer and closer to the AGNs core. However, these models
also predict that at a particular frequency, jet emissions cease to be
dominated by optically--thick synchrotron emission because of
electrons cooling effects, and become dominated by synchrotron
emission in optically--thin regime. Spectra then change from ``flat''
to ``steep''.

Let us review in more detail the \citet{kon81} model. In this
model, the magnetic field $H$ and the distribution of relativistic
electrons (described by a power law
$n(\gamma)=K\gamma^{-(1-2\alpha)}$, where $\gamma$ is the Lorentz
factor) are considered inhomogeneous and scale with the jet
radius as $H(r)\propto r^{-m}$ and $K(r)\propto r^{-n}$. We assume
$m=1$ and $n=2$, which corresponds to assuming the magnetic and the
electron energy densities are in equipartition \citep{bla79}.

For the wavelengths range we are interested in, the synchrotron
spectra from a portion of a relativistic conical jet at a distance
$r$ from the AGN core show two characteristic frequencies:
\begin{itemize}

\item $\nu_{sm}$, the frequency where the synchrotron spectrum has the
  maximum and is estimated by setting the optical depth equal to
  unity. Below $\nu_{sm}$ the emitting source is optically thick to
  synchrotron self-absorption and the flux density rises as
  $\nu^{2.5}$, whereas at $\nu>\nu_{sm}$ the emission has the
  optically--thin synchrotron spectrum $S\propto\nu^{\alpha}$ with
  $\alpha<-0.5$. Under the previous hypothesis, the maximum frequency scales
  as $\nu_{sm}\propto r^{-1}$. It means that the size of the optically
  thick ``core'' of the jet decreases as the observed frequency
  increases.

\item $\nu_{sb}$ ($>\nu_{sm}$), above which spectra steepen owing to
  synchrotron--radiation losses \citep{bla79}. The cutoff occurs when
  the synchrotron cooling time equals the reacceleration time of
  electrons. Contrary to $\nu_{sm}$, $\nu_{sb}$ increases as the
  distance from the core increases, proportionally to $r$. This
  implies that in an AGN jet there is a radius $r_M$ at which
  $\nu_{sm}(r_M)=\nu_{sb}(r_M)$. Thus, $r_M$ is the smallest radius
  from which optically--thin synchrotron emission can be observed with
  its characteristic "steep" spectral index, $\alpha<-0.5$.

\end{itemize}

As said before, the flux density observed from an unresolved conical
jet is given by the sum of the emission from the different portions of
the jet. The flat part of synchrotron spectra is observed up to the
frequency $\nu_M$ (that we call ``break frequency''), i.e. the
frequency at which $\nu_{sm}(r_M)=\nu_{sb}(r_M)$. At frequencies
$\nu>\nu_M$, contributions from regions $r<r_M$ inside the jet are
negligible owing to the cooling of high--energy electrons. The observed
flux density is thus dominated by optically--thin emission from radii
$r\ge(\nu/\nu_M)r_M$ that verify $\nu_{sb}(r)>\nu$ \citep{kon81}.

Therefore, spectra at cm/mm wavelengths can be approximated by two
power laws
\begin{equation}
S(\nu) = \left\{
\begin{array}{ll}
S(\nu_M)\,(\nu/\nu_M)^{\alpha_{fl}} & ~~{\rm if} ~
\nu\le\nu_M \\
S(\nu_M)\,(\nu/\nu_M)^{\alpha_{st}} & ~~{\rm if} ~
\nu\ge\nu_M\,,
\end{array} \right.
\label{e2a}
\end{equation}
where $\alpha_{fl}$ and $\alpha_{st}$ represent the effect of the
nonuniformity of sources in the optically thick and optically thin
regimes.

The frequency $\nu_M$ and the radius $r_M$ are provided by
\citet{bla79} and \citet{kon81} as the function of the relevant physical
quantities of AGN jets. These expressions are quite complex, however,
and depend on too many model parameters. For simplicity (and, e.g., in
agreement with \citealt{mar87} and \citealt{ghi93}), we
assume that the flux density observed at frequency $\simeq\nu_M$
is dominated only by the emission from the region at radius $r_M$, and
that other contributions are negligible. Then, for single emitting
regions, synchrotron spectra can be approximately described by the
homogeneous spherical model, and the break frequency is given by
\citep{pac70}
\beq
\nu_M\propto
S_M^{2/5}\,\theta^{-4/5}\,H^{1/5}\,(1+z)^{1/5}\,\delta^{-1/5}\,,
\label{e2}
\eeq
where $S_M$ is the observed flux density at $\nu_M$ and $\delta$
is the Doppler factor. 
Finally, $\theta$ is the observed angular dimension of the emitting
region: for a narrow conical jet of semiangle $\phi$, the axis of
which makes an angle $\psi$ with the direction of the observer,
$\theta$ can be written as \beq \theta=2{(1+z)^2 \over
D_L}\,r_M\tan(\phi)\,\cos \psi \simeq2{(1+z)^2 \over
D_L}\,r_M\,\phi\,, \label{2eb} \eeq where $D_L$ is the luminosity
distance of the source, and we assume the view angle
$\psi\la10\degr$ and $\cos\psi\simeq1$. As a typical value, we take
$\phi\sim0.1$ \citep{kon81,ghi09}.

Under the hypothesis of equipartition condition, we can also have an
estimate of the intensity of the magnetic field in the emitting
region. Indeed, from the minimum energy conditions \citep{pac70},
the magnetic field intensity is
\beq
H_{\rm eq}=\sqrt{{24 \over 7}\pi u_{\rm min}}\,,
\label{e3}
\eeq
where the minimum energy density $u_{\rm min}$ is
\beq
u_{\rm min}\simeq10^{-23}\Bigg({L \over V}\Bigg)^{4/7}\,,
\label{e4}
\eeq
and $L$ is the luminosity of the emitting component, $V$ the volume in
pc$^3$. In Eq.~(\ref{e4}) we are assuming that the total energy of
electrons is equal to the total energy of protons.

In the literature there is no clear agreement on whether the condition
of equipartition is valid in blazar jets\footnote{According to
  \citet{mar03}, the condition is verified in flat--spectrum quasars,
  but not in BL\,Lacs, where the magnetic field energy is observed to
  be lower than the relativistic particle energy. Indications that the
  brightness temperature of the radio emission in jet components
  rapidly drops to the equipartition limit moving away from the core
  are also given by \citet[][]{lob10} and references therein. On the
  other hand, \citet{ghi10b} argue that blazar jets should be
  matter--dominated on scales where most of their luminosity is
  produced, and sub--equipartition magnetic fields are found only on
  smaller scales.}. The equipartition magnetic field should be
considered only as an upper limit, but because of the weak dependence
of $\nu_M$ on $H$ ($\nu_M\propto H^{1/5}$), we can safely adopt the
equipartition approximation.

The total luminosity in a homogeneous spherical source is related to
the break frequency and to the flux density at $\nu_M$ as (the proof
is given in Appendix\,B) 
\beq 
L=f(\alpha){D_L^2\,\delta^{\alpha-3}
\over
  (1+z)^{1+\alpha}} S_M\nu_M^{-\alpha}\,.
\label{e4a}
\eeq
The factor $f(\alpha)$ depends on the optically--thin spectral index
and on the lower and upper cutoff in the electron distribution
function.

Finally, if the flux density at 5\,GHz is known, $S_M$ can be
extrapolated according to Eq.\,(\ref{e2a}),
$S_M=S_5(\nu_M/5)^{\alpha_{fl}}$. By substituting in Eq.\,(\ref{e2})
the equipartition magnetic field $H_{\rm eq}$ calculated above, the
break frequency $\nu_M$ then only depends on the physical parameters
$z$, $\delta$ and $r_M$, and on the spectral indices before and
after the break frequency, $\nu_M$. Equation\,(\ref{e2}) thus becomes
\beq
\nu_M\approx
C(\alpha,\,\alpha_{fl})[D_L^{\beta_D}(1+z)^{\beta_z}\delta^{\beta_{\delta}}
r_M^{\beta_r}]^{1/\beta}\,, \label{e4b}
\eeq
with $\beta\approx1+0.5\alpha_{fl}$, $\beta_D\approx1$,
$\beta_z\approx-1.5$, $\beta_{\delta}\approx-0.5$
and $\beta_r\approx-1$. The explicit derivation of this equation is
given in Appendix B.

\begin{figure}
\centering
\includegraphics[width=8.cm]{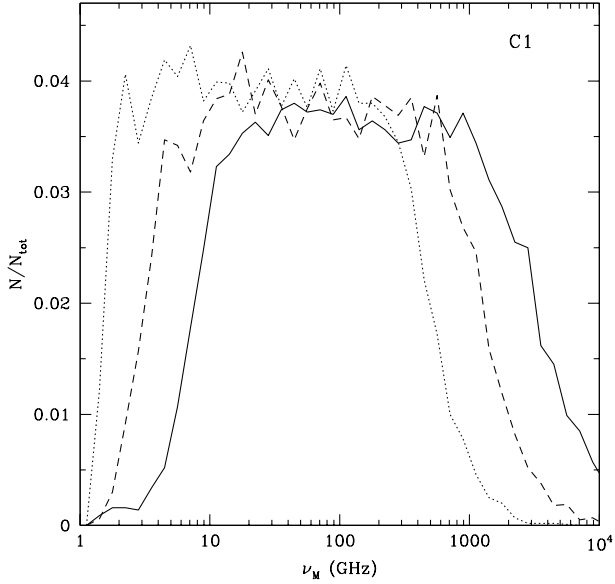}
\caption{Distribution of the break frequency, $\nu_M$, for $10^4$
sources with  5--GHz flux density of 1\,Jy (solid line), 0.1\,Jy
(dashed line) and 10\,mJy (dotted line). In this plot we only
consider $0.01\le r_M\le10$\,pc (case {\bf C1}).}
\label{f6a}
\end{figure}

\begin{figure}
\centering
\includegraphics[width=8.5cm]{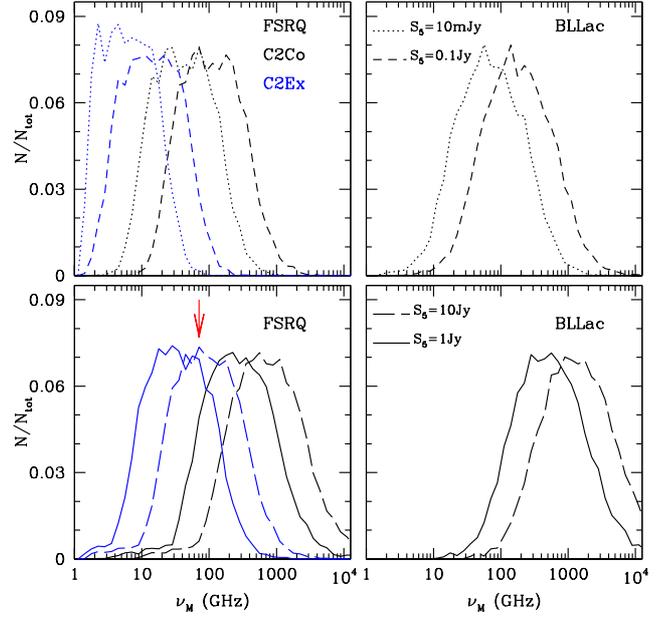}
\caption{Distribution of the break frequency, $\nu_M$, for $10^4$
  sources with 5--GHz flux density of 10/1/0.1/0.01\,Jy (see labels in
  the Figure); the $r_M$ values correspond to the cases {\bf C2Co}
  (black lines) and {\bf C2Ex} (blue lines). In the two upper panels
  only model predictions corresponding to the flux densities of 0.01
  and 0.1 Jy are shown; in the two lower panels, only predictions
  corresponding to 1 and 10 Jy are shown. In the lower panel on the
  left side we also plot the $\nu_M$ (red vertical arrow) of 3C 273
  estimated by \citet{cle83}.} \label{f6b}
\end{figure}

\subsection{Break frequency and magnetic fields in flat--spectrum sources}

The most relevant parameter for the calculation of the break
frequency in blazars spectra is, undoubtedly, the radius $r_M$.

The {\bf radius $r_M$} is the distance from the AGN core of the jet
portion that dominates the emission at $\nu_M$, and it is related to
the break frequency approximately as $\nu_M\propto r_M^{-1}$ (in
agreement with predictions of inhomogeneous models). This parameter
is the most relevant one in the estimate of $\nu_M$ because it
defines the dimension and, thus, the {\it compactness} of the
emitting region at that frequency. At the same time it is also the
most critical one, because of the current uncertainty on its real
value and of the relatively great range of possible values.

In the radio, i.e. synchrotron, regime the AGN jets emission is
expected to be produced at distances along the jet starting from
$10^{-2}$--$10^{-1}$\,pc up to parsec scales
\citep{mar96,lob98,mar03,lob10}. Moreover, the standard models used
to fit the spectral energy distribution (SED) of blazars -- i.e. the
leptonic homogeneous one--zone model \citep[e.g.][]{ghi98,ghi09} --
typically find that the ``dissipation region'' is at a distance from
the black hole of around $10^{-2}$--$10^{-1}$\,pc
\citep{ghi09,ghi10}. However, the parameters of one--zone models are
chosen to fit the high--energy emission of blazars
($\nu\ga10^{14}$\,Hz), whereas they are not able to reproduce the
blazar spectra at cm/mm wavelengths, where the emission is supposed
to come from more external and, thus, larger regions of the
jet. Therefore, we consider the above quoted values as lower limits
for $r_M$. Emission from ultra--compact jet components are also
required for explaining observed spectra that keep flat up to
frequencies of $\ga10^{12}$\,Hz \citep[e.g.,][]{abd10,gon10}. On the
other hand, observed breaks in blazars spectra at few tens of GHz
need higher values of $r_M$. Upper limits for $r_M$ are provided by
VLBI observations, the resolution of which is of fractions of
milliarcseconds at frequencies $\nu\la5$\,GHz \citep[see,
e.g.,][]{sok11}. The typical linear dimensions of the dominant AGN
jet components derived from these observations are of the order of
a few parsecs \citep{ghi93,jia98,lob10}.

To summarize, the range of possible values for $r_M$ should be
$0.01\le r_M\le10\,$pc (i.e., $3\times10^{16}\le
r_M\le3\times10^{19}$\,cm). This is a very large interval and it is
taken as a first working case to calculate the break frequency
(hereafter, case {\bf C1}). In addition, we consider more restricted
intervals of $r_M$ values, distinguishing between FSRQs and BL Lac
objects: we take $0.01\le r_M\le0.3\,$pc for BL\,Lacs and two
different cases for FSRQs, $0.03\le r_M\le1\,$pc (hereafter, {\bf
  C2Co}, i.e. ``more compact'') and $0.3\le r_M\le10\,$pc, a factor of
10 larger (hereafter {\bf C2Ex}, i.e. ``more extended''). $r_M$
values are assumed log--uniformly distributed inside the above
ranges. This separation in these two classes of blazars is based on
the fact that they show different spectral energy distributions
SEDs. On one hand, FSRQs, which are more powerful objects, are
observed to have a synchrotron peak frequency,
$\nu_p$,\footnote{$\nu_p$ represents the frequency at which a
maximum is reached in the synchrotron SED of AGNs (in terms of $\nu
F_{\nu}$).} at lower frequencies, around $10^{12}$--$10^{14}$\,Hz,
without a clear correlation with the radio power. On the other hand,
in BL\,Lacs $\nu_p$ is found to be at higher frequencies and with a
larger range of values, from $10^{13}$ up to $\ga10^{16}$\,Hz
\citep{fos98,abd10}. This difference is motivated by the fact that
BL\,Lacs are characterized by a lower intrinsic power and by a
weaker external radiation field. Consequently, in BL\,Lacs the
cooling through radiation losses is less dramatic and the electrons
can be present with sufficiently high energies to still mantain
synchrotron emission up to these high frequencies \citep{ghi98}. That
is why we assume that BL\,Lacs are in general more compact
objects compared to FSRQs, and that the emitting region at the
break frequency, $\nu_M$, is closer to the AGN core. As a
consequence, the break frequency will be found to be on average at
higher frequencies in BL\,Lacs than in FSRQs.

It is important, however, to stress that there is no physical
clearly established relationship between the break frequency
$\nu_M$ and the peak frequency $\nu_p$ in the synchrotron SED of
AGNs. Their relative position can vary a lot from source to source.
Indeed, although blazar spectra above $\nu_M$ are dominated by
synchrotron emission in the optically--thin regime, the SED of
blazars can still continue to increase up to frequencies $\gg\nu_M$ if
the spectral slope is $-1\la\alpha\la-0.5$. Interesting examples of
SEDs for a complete sample of 104 northern blazars with average
fluxes $S> 1$ Jy at 37 GHz are presented by the \citet{Planck11k}, from
which it is possible to have an idea of the very different positions
of $\nu_M$ and $\nu_p$ in the observed SEDs.

Apart from $r_M$, the other physical quantities relevant for
calculating $\nu_M$ from Eq.\,(\ref{e4b}) are the redshift
distribution, the Doppler factor $\delta$ in AGNs jet, and the
spectral indices before and after the break frequency, $\nu_M$.
Values of these physical quantities are discussed in
Appendix\,\ref{ss3} and in Section\,\ref{s7}.

Figures\,\ref{f6a}--\ref{f6b} show the distribution of break
frequencies in the observer frame for sources with the 5--GHz flux
density of $S_5=10$, 1, 0.1 and 0.01\,Jy, obtained by Eq.\,\ref{e4b}
for the different choices of $r_M$ discussed above. In general, we see
that $\nu_M$ is on average lower for fainter sources, as expected from
the relation $\nu_M\propto S_M^{2.5}$ in Eq.\,(\ref{e2}). for C1 the
large interval of possible values of $r_M$ is reflected in an almost
flat distribution of the break frequencies in the range 20--1000\,GHz,
10--600\,GHz, and 2--200\,GHz for $S_5=1$, 0.1 and 0.01\,Jy,
respectively. A few percent of sources with break frequency values
$\nu_M\ga10^{12}$\,Hz are also found. On the other hand, when
conditions on $r_M$ are stricter, the distributions of $\nu_M$ are
narrower. In the cases {\bf C2Co--C2Ex} we find that most of BL\,Lacs
of 10--1\,Jy have a spectral break at $\nu\gg100$\,GHz, with a
significant fraction at $\nu\ga10^3$\,GHz. Fainter BL Lacs
(100--10\,mJy) have $\nu_M$--distributions peaking around
50--200\,GHz. Therefore, according to our results, BL Lacs should be
observed at cm/mm wavelengths with almost flat spectra, at least for
flux densities $>0.1$\,Jy. The $\nu_M$ distributions of Jy BL Lacs
extend up to $10^{13}$\,Hz, as does the one for {\bf C1}.

For FSRQs, the break frequency varies a lot according to the chosen
range of $r_M$: in bright quasars (10--1\,Jy), $\nu_M$ is mostly
around $10^2$--$10^3$\,GHz if the emitting regions are more compact
and closer to the AGN core (case {\bf C2Co}), and around 10--100\,GHz
in the other case (case {\bf C2Ex}); fainter quasars have $\nu_M\la
400$\,GHz in the case {\bf C2Co} and $\nu_M\la 40$\,GHz in the case
{\bf C2Ex}. In general, the two cases give quite different results for
the value of $\nu_M$. For {\bf C2Co} the majority of
Jy and sub--Jy FSRQs can keep flat spectra up to $\nu\gg 100$\,GHz; on
the contrary, for {\bf C2Ex}, apart from the few
brightest quasars in the sky, the transition to the optically thin
synchrotron spectra can occur already at around 10--20 GHz. As a
comparison, we also plot in Fig.\,\ref{f6b} the {\it break} frequency
value, $\nu_M=60$ GHz, of 3C 273 estimated by multifrequency
quasi-simultaneous observations in \citet{cle83}.  This value is well
within the range expected from the model {\bf C2Ex}. Additional
indications of a break in the spectra of bright blazars around
100\,GHz have been reported by analyses of {\it Planck} data
\citep{Planck11i,Planck11k}.

\begin{figure}
\centering
\includegraphics[width=8.5cm]{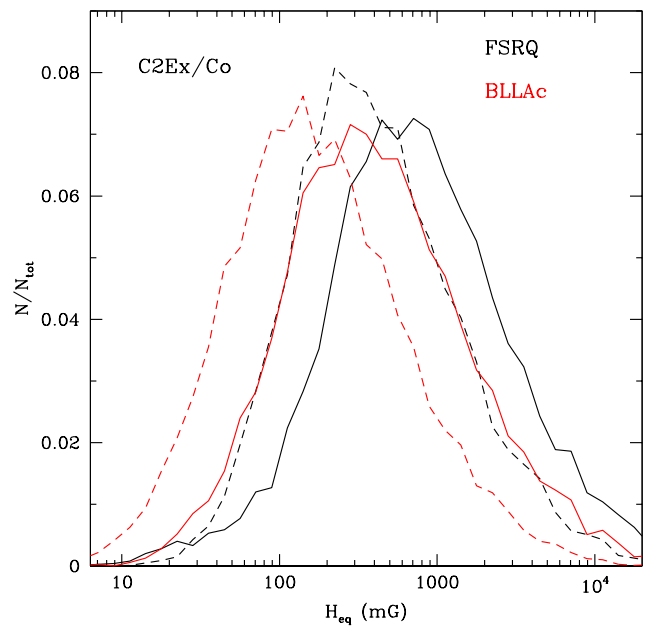}
\caption{Distribution of the magnetic field strength in
  equipartition conditions at a distance of 1\,pc from the AGN
  core. Black curves are for FSRQs and red curves for BL\,Lacs; solid
  curves are for sources with $S_5=1$\,Jy and dashed curves if
  $S_5=0.1$\,Jy.}
\label{f6c}
\end{figure}

Another physical quantity we can derive from the model is the magnetic
field strength in the AGN jet, in equipartition conditions.  We plot
in Fig.\,\ref{f6c} the values of $H_{\rm eq}$ at the jet radius of
1\,pc for FSRQ and BL\,Lac sources with 5--GHz flux densities of
$S_5=1$, and 0.1\,Jy. Magnetic fields in BL\,Lacs are found to be on
average lower than in FSRQs, as expected in less powerful objects.
The distributions for $H_{\rm eq}$ are quite sharp, with peak values
around 200--1000\,mG for FSRQs, and 100--500\,mG for BL Lacs.
Magnetic field strengths of the order of few hundreds of mG are the
values expected using the K$\ddot{o}$nigl model in equipartition
regime for sources of Jy or sub--Jy flux density (e.g., see results in
\citealt{jia98} and \citealt{osu09}).

\section{High--frequency peaked spectrum sources}
\label{s4}

The model described in Sect.\,\ref{s3} does not apply to sources
with inverted spectra ($\alpha\ga 0.3$). Very compact sources have
convex spectra, peaking at GHz frequencies (GHz peaked spectrum,
GPS, sources) or at tens of GHz (high frequency peakers, HFP). It is
now widely agreed that GPS and HFP sources correspond to the early
stages of evolution of powerful radio sources when the radio
emitting region grows and expands in the interstellar medium of the
host galaxy before becoming an extended radio source (see
\citealt{dez05} and references therein; \citealt{ode98} for a
review). At low luminosity, inverted spectra may also correspond to
late stages of AGN evolution, characterized by low
radiation/accretion efficiency \citep{dez10}.

About 5\% of sources in the total NVSS/GB6 sample and $\sim15\%$ of
flat--spectrum sources have inverted spectra between 1 and 5\,GHz;
on the other hand, at high flux densities, these fractions increase
up to about the 10\% and 20\% , respectively (see Table\,\ref{t1}).
The fraction of peaked sources increases significantly in higher
frequency surveys. For example, in the AT20G survey, \citet{mur10}
found that 21\% of sources peak between 5 and 20\,GHz, and 14\% have
a spectrum rising up to 20\,GHz \citep[see also][]{han09}. If we
restrict ourselves to the almost--complete subsample AT20G--d15S100,
we find about 22\% of sources with $\alpha_1^5\ge0.3$, and 18\% with
$\alpha_5^8\ge0.3$, of which about 10\% are inverted up to 20\,GHz.

\begin{figure*}
\centering
\includegraphics[width=7cm]{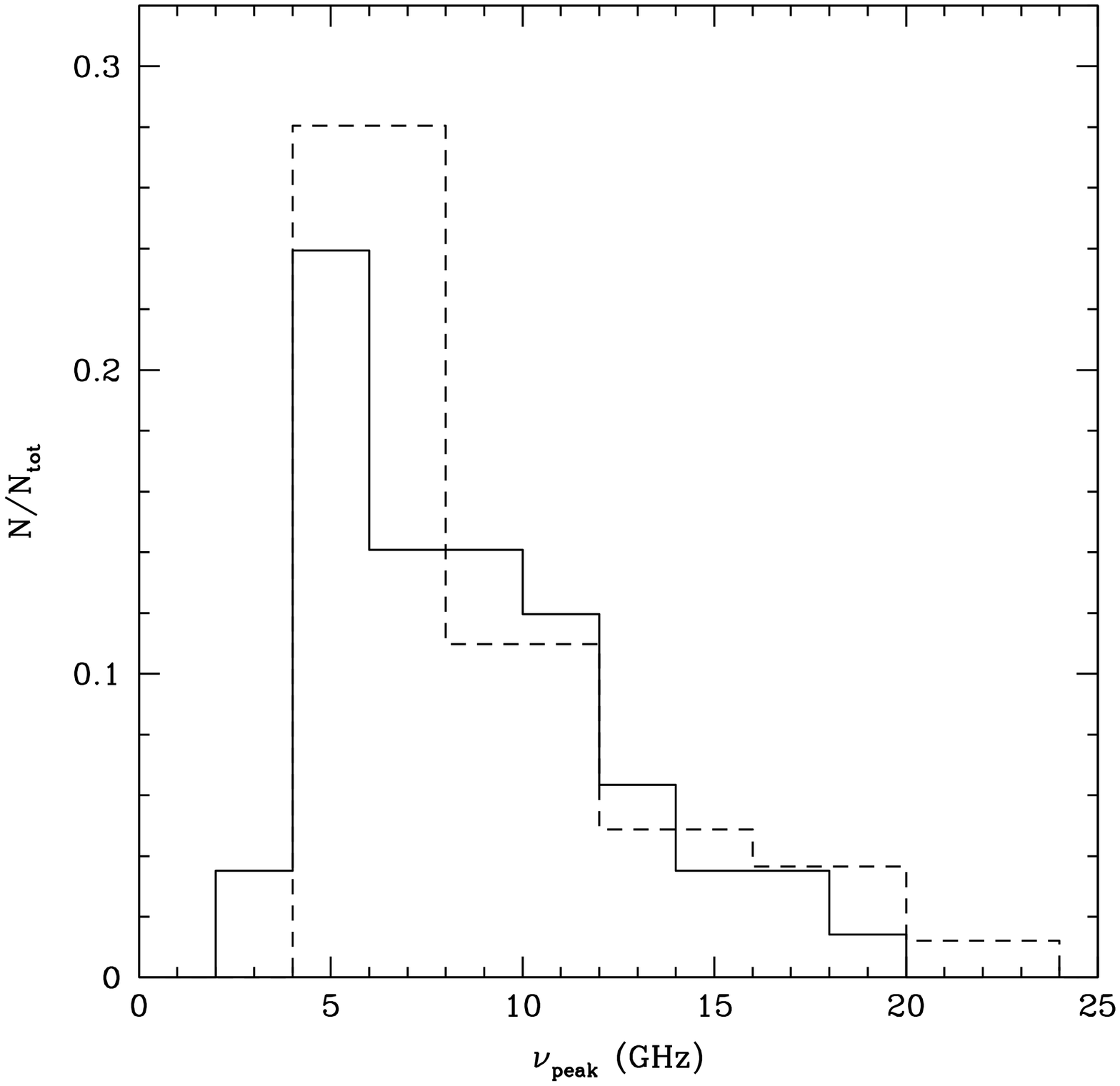}
\includegraphics[width=7cm]{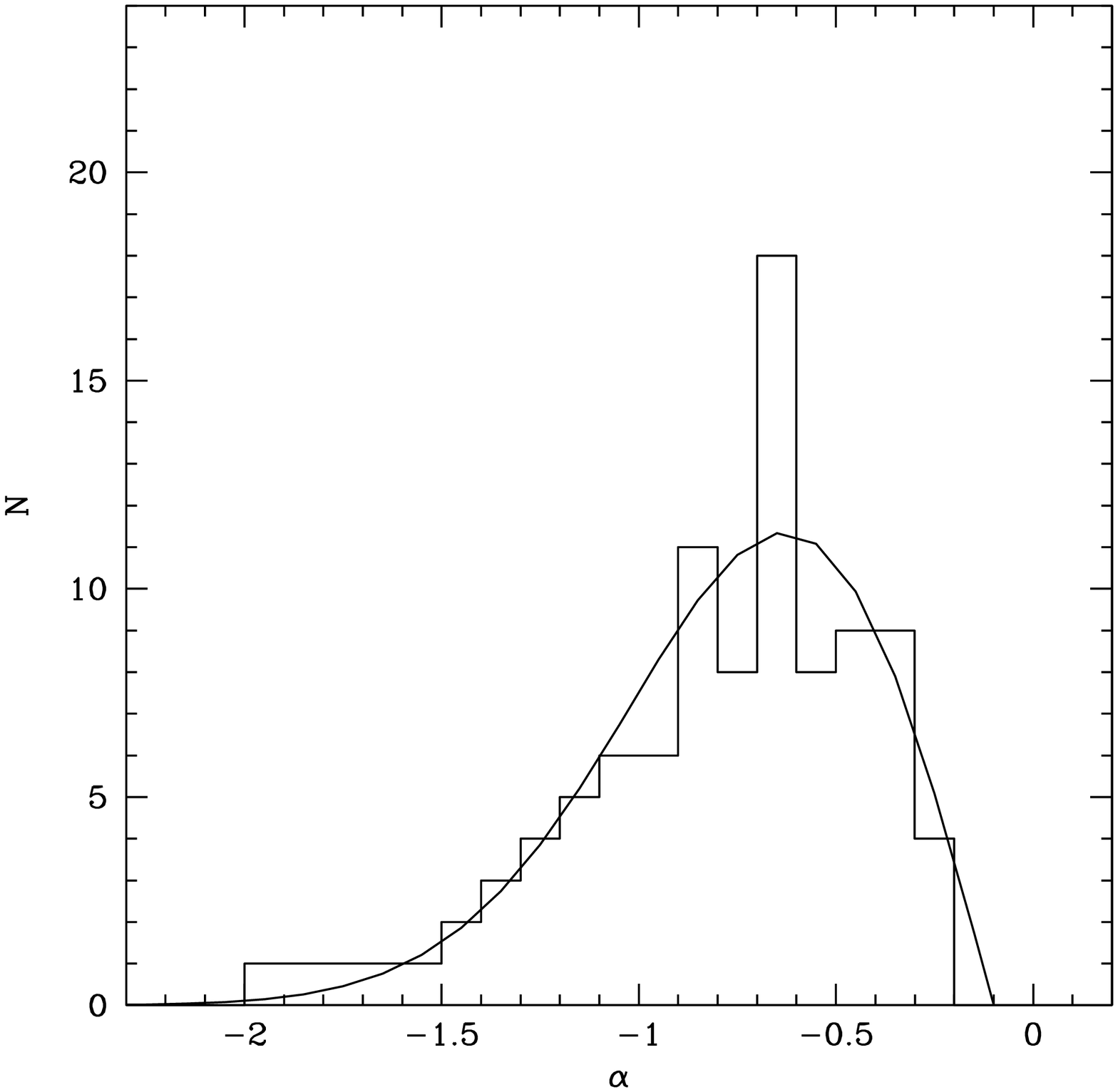}
\caption{{\it Left panel}: distribution of the observed peak
  frequencies for the sample of inverted--spectrum sources defined in
  the text (solid line), and for the GPS/HFP candidates in
  \citet{tor08}. {\it Right panel}: distribution of the spectral
  indexes above the peak frequency for the inverted--spectrum
  sources. The solid line is given by
  $f(\alpha)=-A(\alpha-\alpha_0)\exp[-0.5(\alpha-\alpha_0)^2/\sigma_{\alpha}^2]$,
  with $\alpha_0=-0.10$ and $\sigma_{\alpha}=0.53$.}
\label{f7}
\end{figure*}

Most of these sources cannot be interpreted as truly young GPS/HFPs,
however. Simultaneous multifrequency observations of high-frequency
peaked or inverted-spectrum sources have given evidence of high
flux--density variability or extended emission in a large number of
them, which is not compatible with youth scenario
\citep{dal00,sta03,tor05,tin05,bol06,ori06,Planck11j}, but indicative
that they are most likely blazars caught in an active state, i.e. when
a flaring, strongly self--absorbed synchrotron component dominates the
emission spectrum. Moreover, when observations are not simultaneous,
variability in flat--spectrum sources may lead to misinterpret them as
inverted--spectrum sources.

Our approach to this class of sources is purely statistical, without
dealing with the physical origin of the emission. We are interested in
the spectral shape of inverted--spectrum sources (i.e., all sources
with $\alpha\ge0.3$ between 1 and 5\,GHz) and in statistically
characterizing their behaviour at $\nu>5$\,GHz.

In order to make predictions on the high--frequency number counts, we
need some information on the {\it peak frequency} ($\nu_{\rm peak}$)
\footnote{This peak frequency, characteristic of GPS/HFP sources, must
  be clearly distinguished from the peak frequency, $\nu_p$, in the
  spectra of blazar sources, discussed in Sect. 4.}  and on the
spectral index above the peak frequency.  In the literature we
found three samples of sources with simultaneous measurements of a
wide frequency range, allowing us to recover this information:

\begin{itemize}

\item \citet[][hereafter K99]{kov99} present a nearly simultaneous
  observation of extragalactic radio sources at six frequencies (0.96,
  2.30, 3.90, 7.70, 11.2, and 21.65\,GHz) with the RATAN--600 radio
  telescope. The sample consists of 546 sources selected from the
  \citet{pre85} VLBI survey with a correlated flux density exceeding
  0.1\,Jy at 2.3\,GHz. It includes primarily compact flat--spectrum
  objects because the Preston et al. sample is complete only for
  sources with $\alpha_{2.7}^5\ge-0.5$.

\item \citet[][hereafter D00]{dal00} provide a list of 55 high
  frequency peakers HFPs. This sample was obtained by selecting
  sources with flux density $S_{5GHz}\ge300$\,mJy and
  $\alpha_{1.4}^5>0.5$ from the Green Bank survey 87GB and NVSS. Then
  HFP candidates were observed with the VLA at 1.4, 1.7, 4.4, 5, 8,
  8.5, 15, and 22.5\,GHz, to acquire simultaneous measurements
  of radio spectra and to remove variable sources that do not
  satisfy the criterion $\alpha>0.5$.

\item \citet[][hereafter B04]{bol04} have carried out a follow--up of
  176 sources from the 15--GHz 9C survey. Sources were selected from
  two complete samples with flux limits of 25 and 60\,mJy.
  Simultaneous observations were made at frequencies of 1.4, 4.8, 22,
  and 43\,GHz with the VLA and at 15\,GHz with the Ryle Telescope. In
  addition, 51 sources were also observed at 31\,GHz with the OVRO
  telescope.

\end{itemize}

Spectra of sources from these three samples were first fitted with a
single power law. If the fit was poor, we used a broken power law or a
second--order polynomial in logarithmic space. Then we selected all
sources with $\alpha_1^5=\log(S_5/S_1)/\log(4.8/1.4)\ge0.3$ (where
$S_5$ and $S_1$ are the values of the fit at 4.8 and 1.4\,GHz). In
this way, we created a sample of 142 inverted--spectrum sources: 78 of
them come from K99, 50 from D00 and 30 from B04 (16 are in common
between K99 and D00). Apart from sources of B04, we were able to
estimate the peak frequency only if it occurs at frequencies
$\nu<20$\,GHz. We found that most of spectra peak at $\nu\la10$\,GHz,
and that 117 sources ($\sim80\%$) have $\nu_{\rm peak}<20$\,GHz. In
the B04 sample, where measurements extend up to 43\,GHz, there are two
objects with $20<\nu_{\rm peak}<43$\,GHz and another three that are
still inverted at 43\,GHz. In Fig.\,\ref{f7} we report the
distribution for $\nu_{\rm peak}$ up to 20\,GHz from our sample and
from the GPS/HFP candidates collected by \citet{tor08}: the two
distributions agree well. In the plot we considered
only the 41 sources in the Torniainen et al. sample classified as
``gps'' or ``convex'' and with peak frequency $\nu_{\rm
  peak}\ge4$\,GHz. A similar distribution is also given by
\citet{vol08} for a sample of 91 GPS/HFP candidates (see Fig.\,7 in
that paper).

Results on the distribution of the peak frequency are summarized in
Table\,\ref{t2}, also for frequencies higher than 20\,GHz. We recall
that we select only sources with an inverted spectrum between 1 and
5\,GHz, and so the distribution is not meaningful for GPS sources with
$\nu_{\rm peak}<5$\,GHz. Moreover, in Table\,\ref{t2} we assume that
there are no inverted--spectrum sources with peak at frequencies
higher than 100\,GHz. In any case, current data indicate that the
percentage of sources with $\nu_{\rm peak}>100$\,GHz has to be very
small and is negligible for our purposes. This hypothesis is supported
by the spectral behaviour of sources from simultaneous observations at
20 and 95\,GHz by ATCA \citep{sad08} and by WMAP \citep{gon08}.
Finally, in simultaneous
observations of GPS/HFP candidates by \citet{tor05,tor07,tor08} the
maximum observed peak frequency is $\sim46\,$GHz, even if their
measurements extend up to 250\,GHz.

For 98 objects of our sample of inverted--spectrum sources we can
provide a reliable estimate of the spectral index after the peak
($\alpha_{hi}$). The spectral index distribution, plotted in
Fig.\,\ref{f7}, peaks at $\alpha_{hi}\simeq-0.7$, but it also has a
large tail of steep values up to $-2$. Similar results were obtained
by \citet{sne98}. On the other hand, \citet{dev97} found an average
$\alpha_{hi}\sim-0.7$ but no sources with $\alpha_{hi}<-1$ on a
sample of 72 GPS. Since very few data are currently available on
this subject, we use the spectral index distribution of
Fig.\,\ref{f7} to predict the spectral behaviour of
inverted--spectrum sources ($\alpha>0.3$) at frequencies above
$\nu_{\rm peak}$.

\begin{table}
\caption{Distribution of the peak frequencies for sources with
  $\alpha_1^5 > 0.3$.}
\centering
\begin{tabular}{cc}
\hline
 Peak Frequency (GHz) & Source Fraction \\
\hline
$\nu_{\rm peak}\le6$ & 0.27 \\
$6<\nu_{\rm peak}\le12$ & 0.40 \\
$12<\nu_{\rm peak}\le20$ & 0.15 \\
$20<\nu_{\rm peak}\le40$ & 0.15 \\
$40<\nu_{\rm peak}\le100$ & 0.03 \\
\hline
\end{tabular}
\label{t2}
\end{table}


\section{Extrapolation of 5--GHz number counts to higher frequencies}
\label{s7}

\begin{figure*}
\centering
\includegraphics[width=8.5cm]{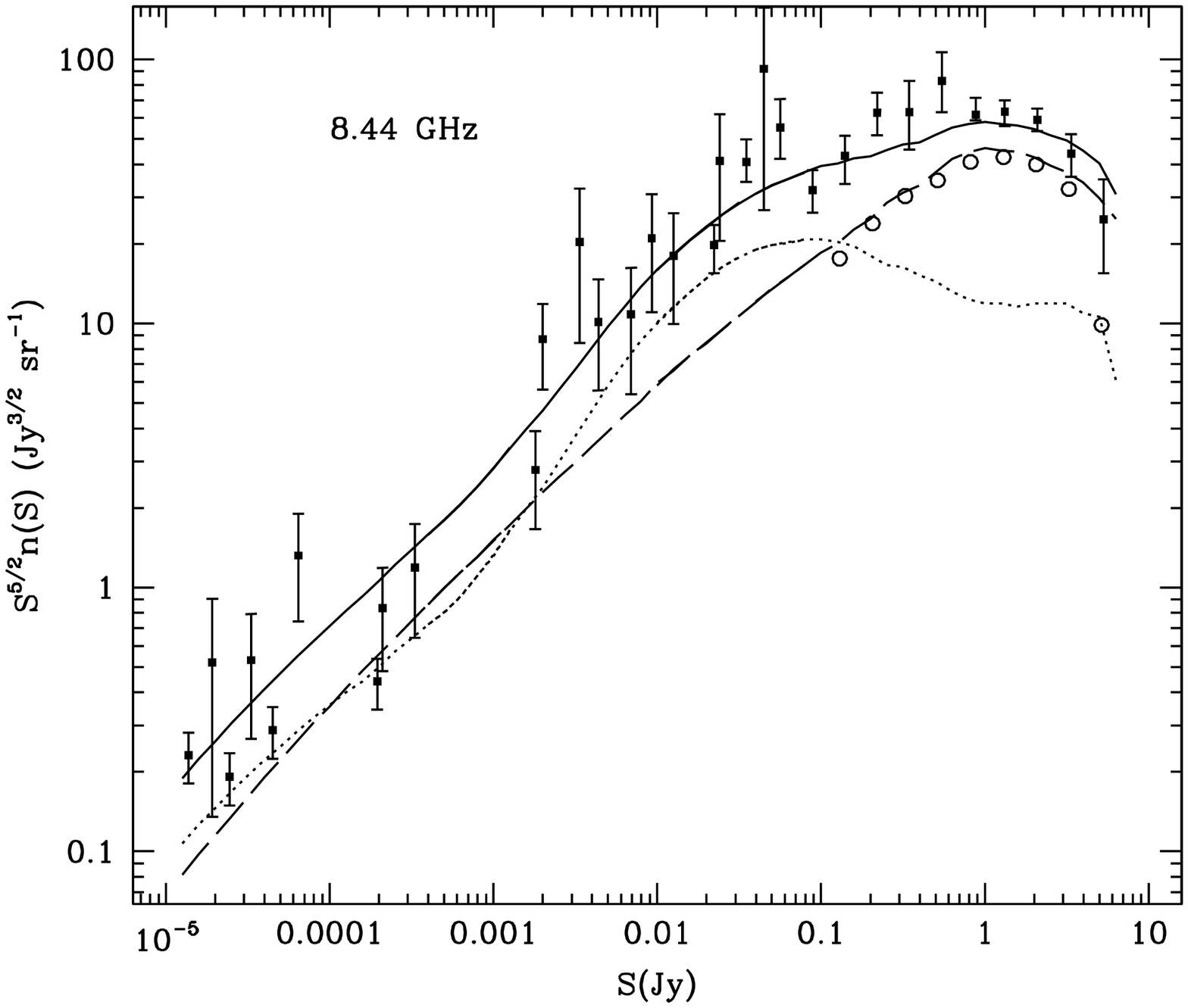}
\includegraphics[width=8.5cm]{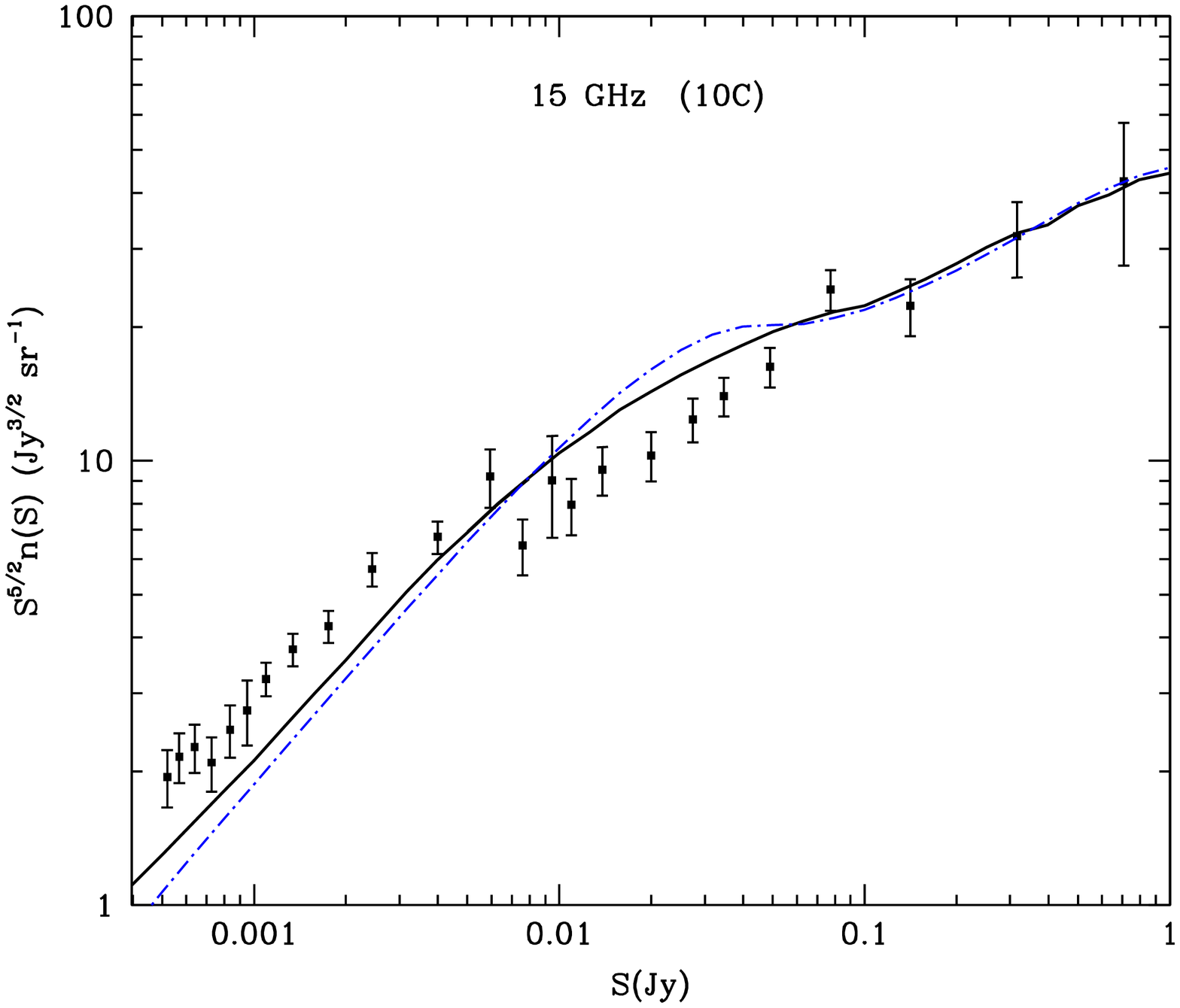}
\caption{Predicted differential number counts normalized to $S^{-2.5}$
  at 8.44\,GHz ({\it Left panel}) and at 15\,GHz ({\it Right panel})
  for the {\bf C1} model (thick continuous line), i.e. the model with
  no difference between the $r_M$ value adopted for FSRQs and BL\,Lac
  objects (see text). At 8.44\,GHz data points are a collection from
  different samples \citep[see][]{dez10}, whereas at 15\,GHz they have
  been computed from the 10C survey. In the {\it left panel}
  predictions for number counts of flat-- (short--dashed line) and
  steep--spectrum (dotted line) sources are also shown, compared with
  results from measurements at 8.4 GHz of flat--spectrum sources by
  the CRATES program (empty circular points). Predictions at 15\,GHz
  from the \citet{dez05} model are also shown (blue dash--dotted
  line).}
\label{f8}
\end{figure*}

\begin{figure*}
\centering
\includegraphics[width=8.5cm]{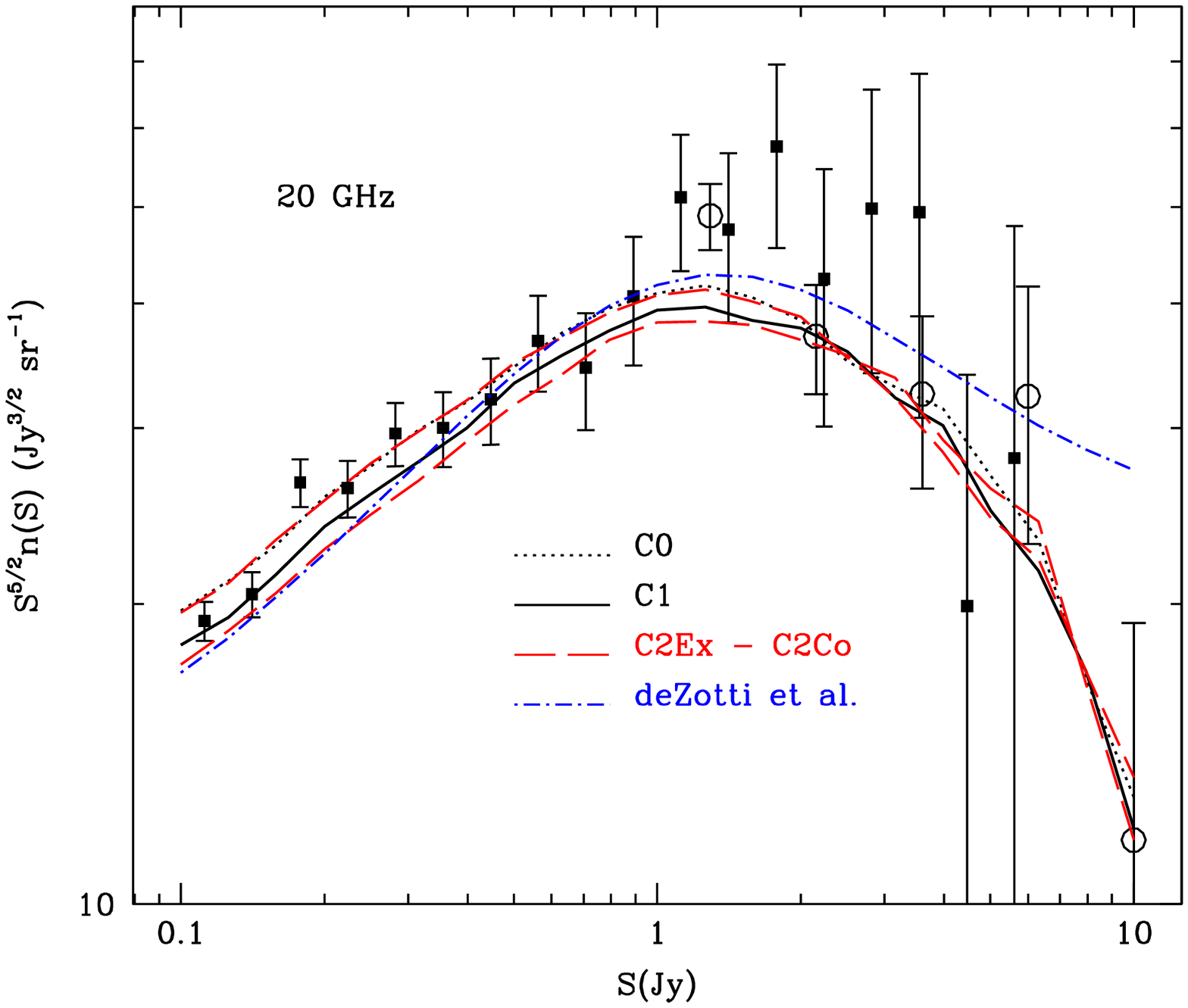}
\includegraphics[width=8.5cm]{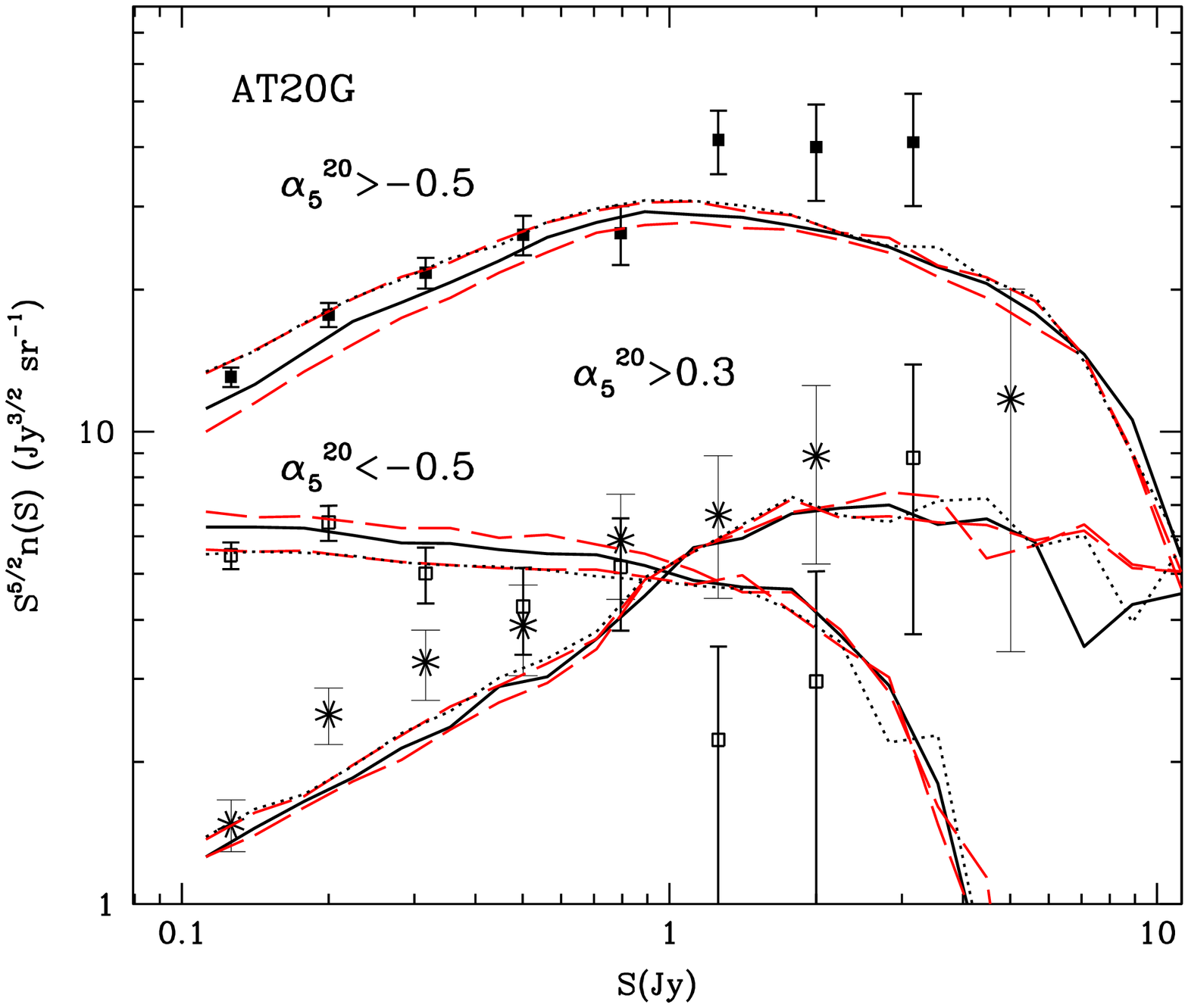}
\caption{Normalized differential number counts at 20\,GHz from
  AT20G--d15S100 sub--sample of the AT20G survey (filled squares; see
  text) compared to predictions of the models described in the text:
  {\bf C0}(dotted lines), {\bf C1} (thick continuous lines), {\bf
    C2Ex} (lower red long--dashed lines) and {\bf C2Co} (upper red
  long--dashed lines) and the \citet{dez05} model (blue dash--dotted
  line). {\it Left panel}: the differential number counts for all
  source populations in the sample. The empty circles are from the
  23--GHz WMAP sources catalogue. {\it Right panel}: differential
  number counts for AT20G-d15S100 sources with
  $-0.5\le\alpha_5^{20}<0.3$ (filled squares), $\alpha_5^{20}<-0.5$
  (empty squares) and $\alpha_5^{20}>0.3$ (star points) and the
  corresponding model predictions.}
\label{f9}
\end{figure*}

Based on the 5~GHz number counts and on information of spectral
properties of ERS described in previous sections, we are now able to
make predictions of number counts at cm/mm wavelengths. We deal with
the complexity of source spectra as follows.

\begin{itemize}

\item We consider three different populations
  of radio sources, according to their spectral index in the frequency
  range 1--5\,GHz: steep--spectrum if $\alpha_1^5<-0.5$;
  flat--spectrum if $-0.5\le\alpha_1^5<0.3$; inverted--spectrum if
  $\alpha_1^5\ge0.3$. Two flat--spectrum populations are taken
  into account: FSRQs and BL\,Lacs.

\item A simulated source catalogue is produced at 5\,GHz. For flat
  plus inverted--spectrum sources we use the $n(S)$ obtained from the
  fit of observational data given by Eq.\,(\ref{e1}). The $n(S)$ of
  steep--spectrum sources is then extracted as the difference between
  the total number counts from the \citet{tof98} model and
  the fit of Eq.\,(\ref{e1}); see also Fig.\,\ref{f2}. The fraction of
  BL\,Lacs among flat--spectrum sources at 5\,GHz, which depends on the
  flux density, is taken from the evolutionary model of \citet{dez05}.

The spectrum for each source is approximated as a power law at
$\nu<5$\,GHz. Spectral indices $\alpha_1^5$ are supposed to follow
the flux--density dependent spectral index distributions obtained from
the NVSS/GB6 sample (see Section\,\ref{ss1} and Table\,\ref{t1}).

\item The 5--GHz flux density of each source is then extrapolated to
  $\nu>5$\,GHz using the following spectral models.

\smallskip\noindent For {\bf steep--spectrum sources} we take
\begin{equation}
S(\nu) = S_{5{\rm GHz}}(\nu/5)^{\alpha_{hi}}\,,
\label{e5a}
\end{equation}
where $\alpha_{hi}=\alpha_1^5-\Delta\alpha$. The spectral steepening
$\Delta\alpha$ is randomly extracted from a Gaussian distribution with
average $0.3$ and dispersion $0.2$, in agreement with the steepening
found in the AT20G--d15S50 sample and shown in Fig.\,\ref{f5}. A small
percentage of flattening or upturning sources is also included.

\smallskip\noindent For {\bf inverted--spectrum sources} we use
\begin{equation}
S(\nu)=S_0(\nu/\nu_0)^k\bigg(1-e^{-(\nu/\nu_0)^{l-k}}\bigg)\,,
\end{equation}
where $\nu_0$ is directly related to the peak frequency in the
spectrum and $S_0=(1-e^{-1})S(\nu_0)$. The coefficients $k$ and $l$
are the spectral indices for the rising and declining parts of the
spectrum. We set $k=\alpha_1^5$, whereas $l$ is extracted from the
best--fit distribution shown in Fig.\,\ref{f7}. The peak frequency
is extracted from the distribution in Table\,\ref{t2}.

\smallskip\noindent For {\bf flat--spectrum sources} we use different
spectral models. In the simplest one ({\bf C0}) radio spectra at
$\nu>5\,$GHz are modelled as power laws with spectral index
$\alpha_{fl}=\alpha_1^5-\Delta\alpha$, where $\Delta\alpha$ is
extracted from Gaussian distributions with
\begin{eqnarray}
<\Delta\alpha>=0.1 ~~{\rm and}~~ \sigma=0.1 & & ~~{\rm
  if}~~~-0.5\le\alpha_1^5<-0.1 \nonumber \\
<\Delta\alpha>=0.2 ~~{\rm and}~~
\sigma=0.2 & & ~~{\rm if}~~~-0.1\le\alpha_1^5<0.3\,. \label{eq:delta}
\end{eqnarray}
These values for $<\Delta\alpha>$ (and the associated dispersions) agree
with observations from the AT20G survey (see Section\,\ref{ss2}).

As another step in modelling the spectra of flat--spectrum ERS, we
introduce a ``break frequency'', as discussed in Sec.\,\ref{s3}.
Spectra are now described by
\begin{equation}
S(\nu) = \left\{
\begin{array}{ll}
S_{5{\rm GHz}}(\nu/5)^{\alpha_{fl}} & ~~5\le\nu\le\nu_M \\
S(\nu_M)\,(\nu/\nu_M)^{\alpha_{st}} & ~~\nu\ge\nu_M\,,
\end{array} \right.
\label{e5b}
\end{equation}
where $\nu_M$ is the break frequency, which is interpreted as the
frequency at which the transition from the optically thick to the
optically thin regime of synchrotron radiation occurs. As above,
the spectral index in the optically thick regime at $\nu>5$\,GHz is
$\alpha_{fl}=\alpha_1^5-\Delta\alpha$, with $\Delta\alpha$ extracted
from the Gaussian distributions of Eq.\,(\ref{eq:delta}), whereas
$\alpha_{st}$ is the spectral index in the optically thin regime,
determined by synchrotron or IC energy losses, and is taken to be
$\alpha_{st}=-0.8\pm0.2$, in agreement with \citet{kel66}.

The break frequency is related to the value of the parameter $r_M$,
discussed in Sec.\,\ref{sect:break}. We considered three cases:

\begin{itemize}

\item {\bf case C1}: $0.01\le r_M\le 10$\,pc, with no difference
between FSRQs and BL Lac objects;

\item {\bf case C2Co}: $0.01\le r_M\le0.3$\,pc for BL Lacs and
$0.03\le r_M\le1$\,pc for FSRQs, i.e. the radio emission in FSRQs
arises from a {\bf ``more compact''} region than in the following
case, {\bf C2Ex};

\item {\bf case C2Ex}: $0.01\le r_M\le0.3$\,pc, the same value as
before for BL Lac objects, and $0.3\le r_M\le10$\,pc for FSRQs, i.e.
the radio emission in FSRQs comes from a relatively {\bf ``more
extended''} region than in the {\bf case C2Co}.

\end{itemize}

\item In the end, simulated catalogues can be extracted at different
  frequencies and for different flux limits. Number counts and
  spectral properties of the three populations of ERS can
  then be compared with observational data.

\end{itemize}


\section{Predictions on number counts and spectral properties of
  ERS at $\nu>5$\,GHz}

\subsection{Number counts: predictions vs observations}

A summary of the relevant data is presented in Appendix\,\ref{s0}.  At
$\nu\la30\,$GHz, we find a general agreement between the observed
$n(S)$ and predictions of all models described in the previous section
({\bf C0}, {\bf C1}, {\bf C2Co} or {\bf C2Ex}). In Fig.\,\ref{f8} we
plot our results at 8.4 and 15\,GHz, but only for our intermediate
model, i.e. {\bf C1}, because of the very small differences among
models. In Fig.\,\ref{f9}, thanks to the almost--simultaneous
measurements at 5 and 20\,GHz for most of the AT20G sources, we test
in more detail the goodness of our predictions for the total number
counts of sources in the sub--sample AT20G--d15S100 (left panel), as
well as for sources within different ranges of $\alpha_5^{20}$ (right
panel). Again, all our models are found to agree very well with the
data. This is an important confirmation that the adopted
number counts at 5\,GHz (especially for flat--spectrum sources) and
spectral index distributions from the NVSS/GB6 sample are essentially
correct.

\begin{figure}
\centering
\includegraphics[width=8.5cm]{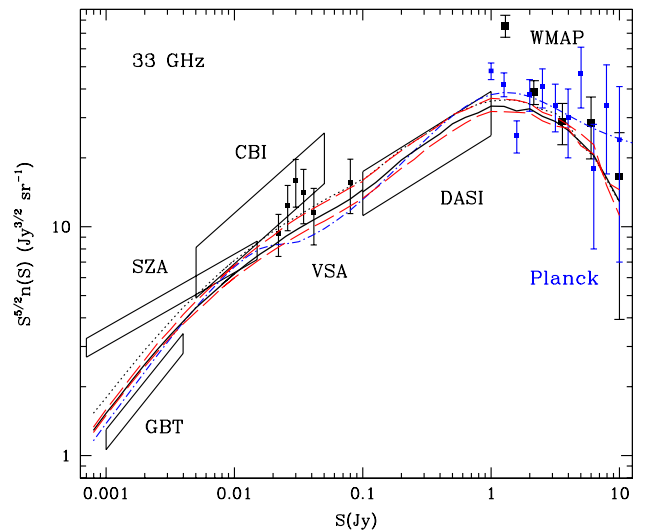}
\caption{Predicted differential number counts at 33\,GHz compared to
  observational data. The lines have the same meaning as in
  Fig.\,\ref{f9}. The area labelled `GBT' shows estimates by
  \citet{maso09}, based on the 31--GHz survey conducted using the GBT
  and OVRO telescopes.}
\label{f10}
\end{figure}

\begin{figure}
\centering
\includegraphics[width=8.5cm]{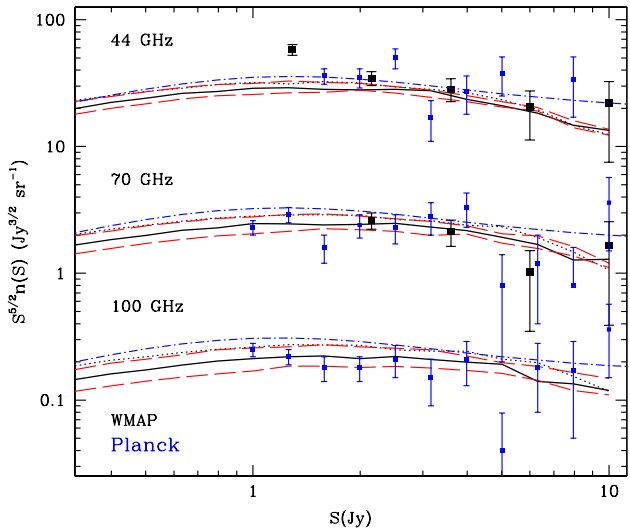}
\caption{Differential number counts predictions as in Fig.\,\ref{f9}
  but at 44, 70 and 100\,GHz. Bigger black points are from the WMAP
  NEWPS catalogue by \citet{mas09} at 41 and 61\,GHz; smaller blue
  points are from the Planck ERCSC. For clarity, number counts
  at 70\,GHz are multiplied by 0.1, at 100\,GHz by 0.01.}
\label{f11}
\end{figure}

The agreement with the data is confirmed at 30--33\,GHz over a very
broad flux density range (Fig.\,\ref{f10}), and it extends up to
100\,GHz, as shown in Fig.\,\ref{f11}.

\begin{figure*}
\centering
\includegraphics[width=8.5cm]{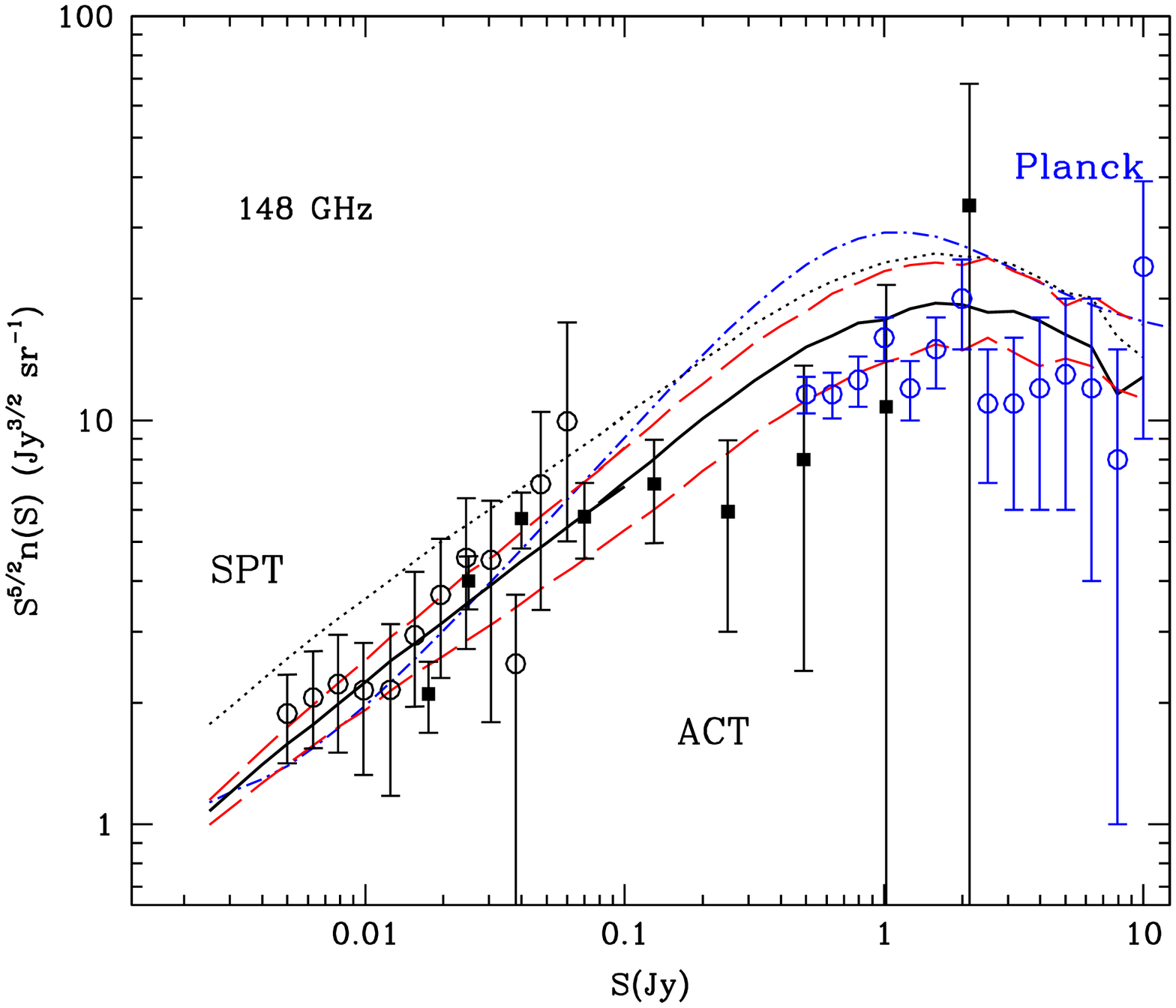}
\includegraphics[width=8.5cm]{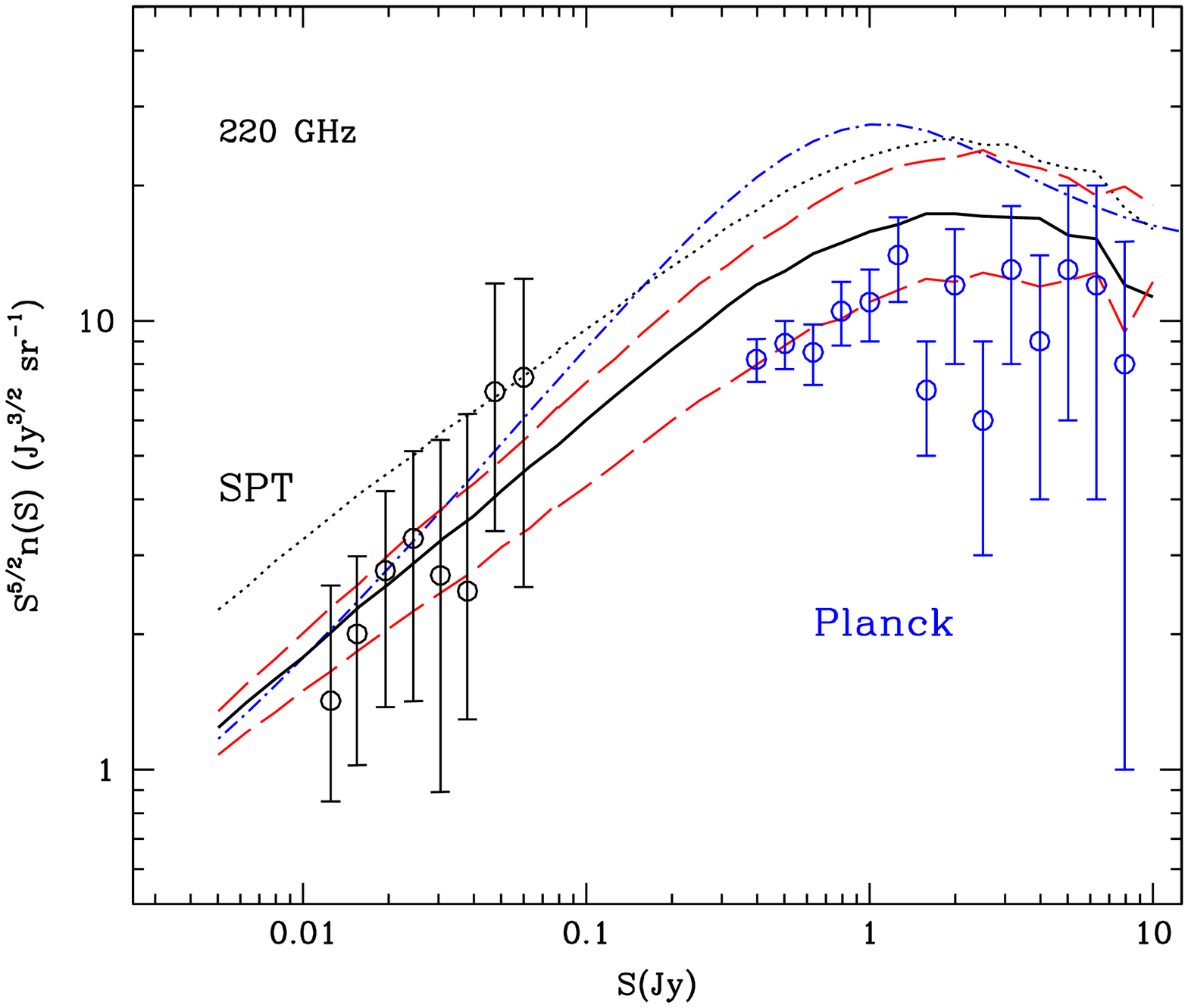}
\caption{Comparison between predicted and observed differential number
  counts at 148\,GHz ({\it left panel}) and at 220\,GHz ({\it right
    panel}). Filled circles: ACT data; open black circles: SPT data;
  open blue circles: {\it Planck} ERCSC counts \citep{Planck11i} at
  143\,GHz (left panel) and 217\,GHz (right panel). The lines have the
  same meaning as in Fig.~\protect\ref{f9}.}
\label{f13}
\end{figure*}

To distinguish among our models we need data at still higher
frequencies ($>100$\,GHz; Fig.\,\ref{f13}). The model without a
spectral break ({\bf C0}) overestimates the counts over essentially
the whole flux--density range. The other models show the largest
differences among each other at the brightest fluxes (at $S=1$\,Jy,
the number counts from the model {\bf C2Co} are almost a factor 2
higher compared to the model {\bf C2Ex}), whereas they tend to
converge at $S<10$\,mJy. A very good match with the data is obtained
with the model {\bf C2Ex}, whereas the other models clearly
overestimate the {\it Planck} counts.

For a direct comparison, we also plot in Fig.\,\ref{f13} the number
counts predicted by the \citet{dez05} cosmological evolution model for
ERS. At bright flux densities ($S>0.1$\,Jy), their results are
compatible with our model {\bf C0}, as expected because both models
use simple power--law approximations for blazar spectra.  However, at
lower fluxes, the differences between these two models increase, and
the De\,Zotti et al. model provides predictions similar to those
obtained from our models including a {\it break frequency}. This is no
surprises because it can be attributed to the different 5--GHz $n(S)$
for flat--spectrum sources that our models use (as shown in
Fig.\,\ref{f2}), if compared to the De Zotti et
al. model.\footnote{The excess of steep--spectrum ERS -- with respect
  to the observed ones -- adopted by the De Zotti et al. model at 5
  GHz can clearly mimic the effect of the break frequency in the
  spectra of flat--spectrum ERS at mm wavelengths at least partially.}

\begin{figure}
\centering
\includegraphics[width=8.5cm]{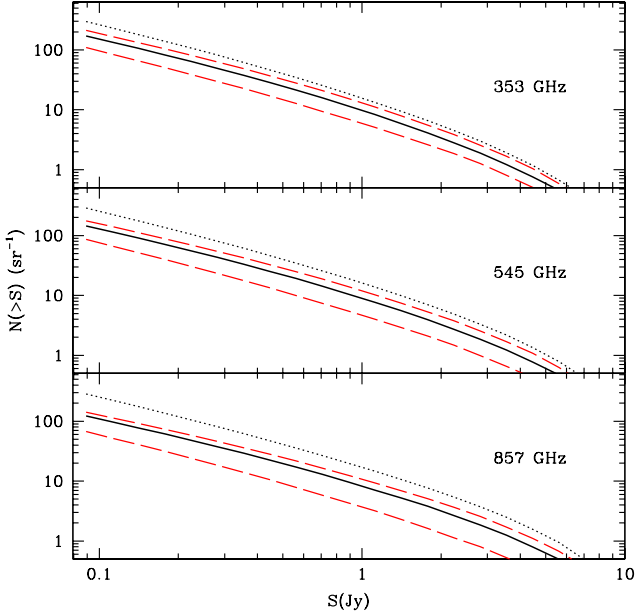}
\caption{Predictions on integral number counts at 353, 545 and
857\,GHz from the models {\bf C0}, {\bf C1}, {\bf C2Co}, {\bf C2Ex}
discussed in Section 7. The plotted lines have the same meaning as
in Fig.\,\ref{f9}} \label{f14}
\end{figure}

 Finally, a clear distinction among our models should be
provided by the radio source counts at the highest {\it Planck} HFI
frequencies. Our predictions on integral number counts, $N(>S)$, are
presented in Fig.\,\ref{f14} and in Table\,\ref{t5}. As an example, the
number of objects brighter than 1\,Jy should be reduced by
$\sim40\%$ in the range 353--857\,GHz for the model {\bf C2Ex}, but
by only $\sim20\%$ or less in the other cases. Moreover, we predict
that the {\it Herschel}--ATLAS survey \citep{eales} will detect from
25 (model {\bf C2Ex}) to 50 (model {\bf C2Co}) blazars brighter than
50\,mJy at 500$\,\mu$m over its area of 550$\deg^2$. For comparison,
the \citep{dez05} model yields $\sim80$ blazars in the same area
\citep{gon10}.

\subsection{Predicted spectral properties of radio sources}

As a further test, we considered the average and the median spectral
indices of ERS in catalogues selected at high frequencies. In the ACT
sample, all but two of the 42 detections with flux density
$S_{148}\ge50\,$mJy have counterparts in AT20G at 5 and 20\,GHz
\citep{mar10}. An average spectral steepening is found above 20\,GHz:
the median spectral index and the dispersion of the distribution are
$\alpha_5^{20}=-0.07\pm 0.37$, $\alpha_5^{148}=-0.20\pm 0.21$ and
$\alpha_{20}^{148}=-0.39\pm 0.24$. Somewhat flatter spectral indices
are observed for SPT sources (57 objects with a 5--$\sigma$ detection
at 148 and 220\,GHz) between 5 and 148\,GHz,
$<\alpha_5^{148}>=-0.13\pm 0.21$, with a substantial steepening
between 148 and 220\,GHz, $<\alpha_{148}^{220}>=-0.5$
\citep{vie09}. To compare these data with model predictions, we
produced simulated catalogues of sources selected at 148\,GHz with the
same flux limits as the SPT and ACT catalogues and with
$S_{20}\ge50$\,mJy, and we computed the mean and median spectral
indices among pairs of frequencies in the range 5--220 GHz; results
are reported in Table\,\ref{t3}.

Similarly, we used the models to simulate samples selected at 30\,GHz
with a completeness limit of 1\,Jy to compare the predicted median
values of spectral indices between 30\,GHz and higher Planck
frequencies and the associated dispersions with the observed values,
given in Table\,2 of the \citet{Planck11i}. The results are shown in
Table\,\ref{t4}.

\begin{table}
\caption{Number of ERS brighter than $1$ and 0.5\,Jy in the
  full sky at the highest {\it Planck} frequencies predicted by our
  models.}
\centering
\begin{tabular}{ccccccccc}
\hline
 & & \multicolumn{3}{c}{$N(\ge1$\,Jy)} & &
\multicolumn{3}{c}{$N(\ge0.5$\,Jy)} \\
\hline
$\nu$[GHz] & & 353 & 545 & 857 & & 353 & 545 & 857 \\
\hline
model & & & & & & & & \\
C0   & & 205 & 207 & 218 & & 507 & 512 & 522 \\
C1   & & 130 & 116 & 107 & & 311 & 278 & 246 \\
C2Co & & 174 & 156 & 141 & & 406 & 361 & 313 \\
{\bf C2Ex} & & {\bf 77} & {\bf 60} & {\bf 47} & & {\bf 184} & {\bf
  143} & {\bf 112} \\
\hline
\end{tabular}
\label{t5}
\end{table}

\begin{table*}
\caption{Means, dispersions, and medians of spectral indices calculated
  between different pairs of frequencies as predicted by our models.}
\centering
\begin{tabular}{cccccccc}
\hline
 \multicolumn{2}{c}{$\nu$[GHz]} & [5,\,20] & [20,\,148] & [5,\,148] &
 & [5,\,148] & [148,\,220] \\
\hline
\hline
 & & \multicolumn{3}{c}{ACT (median $\pm\sigma$)} & &
\multicolumn{2}{c}{SPT (mean $\pm\sigma$)} \\
\hline
            & & $-0.07\pm0.37$ & $-0.39\pm0.24$ & $-0.20\pm0.21$ & &
$-0.13\pm0.21$ & $-0.50$ \\
\hline
\hline
 & model & \multicolumn{3}{c}{ACT simulated sample} & &
\multicolumn{2}{c}{SPT simulated sample} \\
\hline
$<\alpha>\pm\,\sigma_{\alpha}$ & C0 & $-0.08\pm0.30$ & $-0.16\pm0.27$ &
$-0.13\pm0.26$ & & $-0.08\pm0.29$ & $-0.14\pm0.32$ \\
            & C1 & $-0.08\pm0.33$ & $-0.25\pm0.30$ & $-0.18\pm0.28$ & &
$-0.16\pm0.31$ & $-0.37\pm0.38$ \\
            & C2Co & $-0.07\pm0.31$ & $-0.20\pm0.28$ & $-0.15\pm0.27$ & &
$-0.14\pm0.30$ & $-0.36\pm0.39$ \\
            & {\bf C2Ex} & {\bf -0.07\ }$\pm${\bf\ 0.36} & {\bf
  -0.35\ }$\pm${\bf\ 0.31} & {\bf -0.23\ }$\pm${\bf\ 0.30} & &
{\bf -0.22\ }$\pm${\bf\ 0.33} & {\bf -0.50\ }$\pm${\bf\ 0.37} \\
\hline
$\alpha$\,({\rm median}) & C0 & -0.10 & -0.17 & -0.13 & & -0.09 & -0.15 \\
             & C1 & -0.10 & -0.24 & -0.18 & & -0.17 & -0.37 \\
             & C2Co & -0.09 & -0.21 & -0.15 & & -0.15 & -0.37 \\
             & {\bf C2Ex} & {\bf -0.09} & {\bf -0.35} & {\bf -0.24} &
& {\bf -0.24} & {\bf -0.55} \\
\hline
\end{tabular}
\label{t3}
\end{table*}

In general, we see that models including a {\it break frequency} in
blazars spectra are required to explain the median spectral indices
$\la-0.3$ observed when the surveys are performed at frequencies
$\ge100$\,GHz. In particular, the models {\bf C1} and {\bf C2Ex} yield
median spectral indices very similar to the Planck ERCSC ones,
i.e. $-0.4\la\alpha_{30}^{\nu_{\rm HFI}}\la-0.3$. They also yield
median steepenings $\alpha_{5}^{20}-\alpha_{20}^{148}$ of 0.14 ({\bf
  C1}) and 0.26 ({\bf C2Ex}) to be compared with 0.32 from ACT data,
and $\alpha_{5}^{148}-\alpha_{148}^{220}$ of 0.22 and 0.28,
respectively, to be compared with $0.37$ from SPT data (see
Table\,\ref{t3}). Remarkably, models including a spectral break (at
$\nu\la100$\,GHz) give quite ``flat'' mean 5--148\,GHz spectral
indices ($<\alpha_5^{148}>\simeq-0.2$), very close to the value found
by \citet{vie09}. This is because of the 148\,GHz selection that
emphasizes sources with flatter spectra.

On the other hand, the median spectral indices yielded by models {\bf
  C0} and {\bf C2Co} are {\it not} compatible with observations at
frequencies $\ga100$\,GHz.

\begin{table}
\caption{Predicted median spectral indices of ERS between 30\,GHz and
  the {\it Planck} frequency channel indicated in the first row. The
  corresponding median values calculated for the spectral indices of
  ERS in the {\it Planck} ERCSC are taken from the \citet{Planck11i}.}
\centering
\begin{tabular}{ccccccc}
\hline
$\nu$[GHz] & & 44 & 70 & 100 & 143 & 217 \\
\hline
Planck & & -0.06 & -0.18 & -0.28 & -0.39 & -0.37 \\
\hline
model & & & & & & \\
C0   & & -0.11 & -0.13 & -0.14 & -0.14 & -0.14 \\
C1   & & -0.18 & -0.23 & -0.27 & -0.31 & -0.34 \\
C2Co & & -0.13 & -0.17 & -0.19 & -0.21 & -0.24 \\
{\bf C2Ex} & & {\bf -0.22} & {\bf -0.28} & {\bf -0.34} &
{\bf -0.39} & {\bf -0.44} \\
\hline
\end{tabular}
\label{t4}
\end{table}


\section{Conclusions}

The main goal of this paper was to present physically grounded models
to extrapolate the number counts of ERS, observationally determined
over very large flux density intervals at cm wavelengths down to mm
wavelengths, where the majority of the experiments aimed at
accurately measuring CMB anisotropies are carried out. Accurate
extrapolations are necessary to minimize and/or to control the
contamination of CMB maps by unresolved extragalactic sources.
Moreover, this paper makes a first attempt at constraining the most
relevant physical parameters that characterize the emission of
blazar sources by using the number counts and the spectral
properties of ERS estimated from high--frequency radio surveys.

We focussed on spectra of blazars that dominate the mm-wave number
counts of ERS at bright flux densities. In the most recent data sets
\citep[e.g.,][]{mas10c,Planck11i,Planck11k}, a relevant steepening in
blazar spectra with emerging spectral indices in the interval between
$-0.5$ and $-1.2$, is commonly observed. We interpreted this spectral
behavior as caused, at least partially, by the transition from the
optically--thick to the optically--thin regime in the observed
synchrotron emission of AGN jets \citep[see, e.g.,][]{mar96}. Indeed,
a ``break'' in the synchrotron spectrum of blazars above which the
spectrum becomes ```steep'' is predicted by models of synchrotron
emission from inhomogeneous unresolved relativistic jets
\citep{bla79,kon81,mar85}. We estimated the value of the frequency
$\nu_M$ at which the break occurs on the basis of the ERS flux
densities measured at 5\,GHz and of the most typical values for the
relevant physical parameters of AGNs. In the framework of these
models, the most relevant and critical physical parameter is the
dimension of the region (approximated as homogeneous and spherical)
that is mainly responsible of the emission at the break frequency. For
a conical jet model, this parameter can be related to the distance of
the emitting region from the AGN core ($r_M$; see Appendix A).

We have investigated four possible cases. The simpler and basic one,
{\bf C0}, assumes simple power-law spectra above 5\,GHz, with a
dispersion of spectral indices. The other three models feature at
high frequencies a spectral break and make different choices for
the parameter $r_M$. In model {\bf C1} the range of $r_M$ (and
therefore of $\nu_M$) is very broad and should include all most
likely values expected from theoretical models \citep[see, e.g.][for
a recent review]{lob10}. Models {\bf C2Co} and {\bf C2Ex} assume
different distributions of $r_M$ for BL\,Lacs and FSRQs, with the
former objects that generate, in general, the synchrotron
emission from more compact regions, implying higher values of
$\nu_M$ (above 100\,GHz for bright objects). These two models differ
only in the $r_M$ distributions for FSRQs: in the {\bf
  C2Co} model the emitting regions are more compact, implying values of
$\nu_M$ partially overlapping with those for BL\,Lacs, whereas in
the {\bf C2Ex} model they are more extended and the values of
$\nu_M$ are consequently lower and clearly
distinguishable from those adopted for BL\,Lacs (approximately in
the range of 10--300\,GHz for $S\ga1$\,Jy).

Obviously, high frequency ($\nu \ge 100\,$GHz) data are the most
powerful for distinguishing among those models. These data clearly
require spectral breaks, thus ruling out model {\bf C0}, in spite of
the average steepening that was introduced in the spectral
indices of ERS at $\nu>5\,$GHz, and they favour the model {\bf
  C2Ex}. According to this, most of the FSRQs (which are the dominant
population at low frequencies and at Jy flux densities), differently
from BL\,Lacs, should bend their flat spectrum before or around
100\,GHz. The model also predicts a substantial increase of the BL\,Lac
fraction at high frequencies and bright flux densities. This is indeed
observed: a clear dichotomy between FSRQs and BL\,Lac objects has been
found in the {\it Planck} ERCSC, where almost all radio sources show
very flat spectral indices at LFI frequencies,
i.e. $\alpha_{LFI}\ga-0.2$, whereas at HFI frequencies, BL\,Lacs keep
flat spectra, i.e. $\alpha_{HFI}\ga-0.5$, and a high fraction of
FSRQs show steeper spectra, i.e. $\alpha_{HFI}<-0.5$. Moreover, the
fraction of BL\,Lacs above 1\,Jy increases from 10\% in the 5--GHz
\citet{kuh81} sample to 20\% in the 37\,GHz selected sample discussed
by the \citet{Planck11k}. More constraints are expected from sub-mm
counts of blazars produced by the Herschel surveys \citep{gon10}.

According to our model, these results imply that the parameter $r_M$
should be of parsec--scales, at least for FSRQs, in agreement with
theoretical predictions \citep{mar85}, whereas values of $r_M\ll1\,$pc
should be only typical of BL\,Lac objects or of rare quasar sources.

On the other hand, the model {\bf C2Ex} slightly underestimates the
number counts at 20--30\,GHz and predicts too steep median spectra at
the {\it Planck} LFI frequencies. This indicates that spectral breaks
at $\nu < 20$\,GHz are rarer, i.e. the emitting regions of FSRQs,
particularly of the fainter ones, are somewhat more compact than
implied by this model. The possibility of more compact AGN cores
associated with less powerful sources could agree better
with physical models of the AGN jet emission \citep[e.g.,][]{ghi98},
and could provide an improvement of the model predictions at tens of
GHz.

The physical model used in this paper adopts only a simplified
description of the synchrotron emission in AGN jets and, of course,
does not pretend to take into account the complexity of all
physical mechanisms involved. Nevertheless, it is capable of providing
a clear interpretation of several features observed in blazar
spectra. The very interesting quantitative results obtained on number
counts and on spectral index distributions of ERS in the whole
frequency interval from 5 to 220\,GHz clearly support this
conclusion. A step forward will be the re-analysis of the luminosity
functions of blazar sources at radio frequencies, to follow the
cosmological evolution of this class of sources for improving the fit,
and the corresponding interpretation of the source number counts at
$\nu > 100$\,GHz and at faint flux levels. However, this further step
will be the subject of a following paper.

\begin{acknowledgements}

MT would like to thank Monica Orienti for the useful and
enlightening discussions about the main physical mechanisms in AGNs.
MT thanks the Physikalisch--Meteorologisches Observatorium
Davos/World Radiation Center (PMOD/WRC) for having provided him with
facilities to carry out the present research. LT and EMG acknowledge
partial financial support from the Spanish Ministry of Education and
Science (MEC), under project AYA2007--68058--C03-02, and from the
Spanish Ministry of Science and Innovation (MICINN), under project
AYA2010--21766--C03-01. GDZ acknowledges financial support from ASI
(ASI/INAF Agreement I/072/09/0 for the Planck LFI activity of Phase
E2).

\end{acknowledgements}


\appendix


\section{Large--area surveys of radio sources}
\label{s0}

\subsection{Low--frequeny data: the catalogues of steep-- and
  flat--spectrum sources}

Large--area deep surveys are specially present at GHz frequencies
and have allowed us to make statistical studies of the spectral
behaviour of ERS for the basic classification into steep-- and
flat--spectrum sources.

\begin{itemize}

\item By using the NRAO VLA Sky Survey \citep[NVSS;][]{con98} at
  1.4\,GHz and the Green Bank survey \citep[GB6;][]{gre96} at
  4.85\,GHz, we built a large and complete catalogue of steep-- and
  flat--spectrum sources at 5\,GHz. These two surveys overlap at
  declinations of $0^{\circ}<\delta<-75^{\circ}$ (i.e., an area
  $\Omega\simeq6.07$\,sr) and have allowed us to calculate the
  spectral indices of ERS in almost half of the sky. The resolution of
  the two surveys is quite different, being 45\,arcsec in NVSS and
  3.5\,arcmin in GB6, whereas the flux limits are 2.5 and 18\,mJy
  respectively.

At first, we decided to consider only GB6 sources with flux density
$S_{4.85}\ge100$\,mJy and Galactic latitude $|b|\ge10^{\circ}$.  The
100--mJy flux limit for GB6 sources is well above the flux limit of
the two surveys and guarantees that we are not losing sources with a
rising spectrum, $\alpha>0$. Moreover, it minimizes the effects of
flux density errors. The Galactic cut at latitude $<10^{\circ}$ is
to avoid Galactic sources in the sample. In addition, we excluded GB6
sources flagged as ``W'' (not reliable source) and ``C'' (confused
source). We cross--matched the GB6 source positions with the NVSS
catalogue by taking all positive matches within a position offset
of 89\,arcsec. We used a slightly bigger maximum position offset
with respect to \citet{hea07} for reducing the number of GB6 sources
without NVSS counterparts. In this way, we found 8127 sources with
a counterpart in NVSS and only 23 without. Finally, whenever more than
one of the NVSS sources fell within the GB6 beam, as a consequence of the
better NVSS angular resolution, which may lead to individually
resolved multiple components, we summed their fluxes, correcting for
the effect of the GB6 beam. We ended up with 2975 flat--spectrum
sources (corresponding to $\sim37\%$ of the total number of sources
in the sample) and 5152 steep--spectrum sources ($\sim63\%$).
Hereafter, we indicate this source catalogue with spectral
information as our {\bf NVSS/GB6} sample.

\item The CRATES programme \citep{hea07} has also carried out an almost
  all--sky sample of flat--spectrum sources brighter than 65\,mJy at
  5\,GHz, using the existing surveys at GHz frequencies. Then they
  assembled the 8.4--GHz flux densities of flat--spectrum sources from
  observations done by CLASS \citep{mye03} and from new observations
  by VLA and ATCA. To mantain uniformity in our analysis, we took
  into account only CRATES sources in the GB6 area, i.e. where spectra
  are computed from NVSS and GB6 measurements. In this area, the authors
  found $\sim5000$ flat--spectrum sources with 5--GHz flux density
  $S\ge65$\,mJy (about a 33\% of total sources).

\item A sample of flat--spectrum sources is also provided by the
  Parkes quarter--Jy sample (\citealt{jac02}; see also
  \citealt{dez05}). It consists of 878 objects selected at 2.7\,GHz
  from several complete surveys of the Parkes radio source catalogue,
  and having spectral index between 2.7 and 5\,GHz
  $\alpha_{2.7}^{5}>-0.4$. The flux limit of these surveys varies
  between 0.1 and 0.6\,Jy, although it is 0.25\,Jy for most of them.

\end{itemize}

\subsection{Surveys at frequencies higher than 10\,GHz}

Recent experiments have surveyed large areas of the sky at frequencies
higher than 10\,GHz. They are important to test the validity of our
predictions on number counts of ERS, but also to provide
more direct information on the spectral shape of sources at cm/mm
wavelengths.

\begin{itemize}

\item The Ryle--Telescope 9C surveys \citep{tay01,wal03} have provided
  a catalogue of sources at 15\,GHz with a completeness limit of
  25\,mJy. These surveys cover the fields observed by the Very Small
  Array (VSA), corresponding to an area of
  $\sim520\,$deg$^2$. Moreover, \citet{wal09} have reported on a
  series of deeper regions, amounting to an area of 115\,deg$^2$
  complete to approximately 10\,mJy, and of 29\,deg$^2$ complete to
  approximately 5.5\,mJy. Finally, the Tenth Cambridge (10C) Survey
  \citep{ami10} has covered an area of $\approx27$\,deg$^2$ at 15.7 GHz down
  to a completeness limit of 1\,mJy (within it, some deeper areas, covering
  $\approx12$\,deg$^2$, are complete down to 0.5\,mJy).

\item A 20--GHz survey of the whole southern sky has been carried out
  by the Australian Telescope Compact Array (ATCA) from 2004 to
  2008. The full source catalogue (AT20G) is presented in
  \citet{mur10} and in \citet{mas10b}, and it includes 5890 sources
  above a flux--density limit of 40\,mJy. The completeness of the
  AT20G catalogue is 93 per cent above 100\,mJy and 78 per cent above
  50\,mJy in regions south of declination $-15^{\circ}$. Most of
  sources with declination $\delta<-15^{\circ}$ were also followed--up
  at 5 and 8\,GHz by near--simultaneous observations.

In our analysis we made an extensive use of this survey but we
limited ourselves to the almost--complete sample of the AT20G
catalogue at declinations $\delta<-15^{\circ}$. It consists of 2195
sources with flux density $S\ge100$\,mJy and of 1612 with $50\le
S<100\,$mJy. We called these two sub--samples {\bf
AT20G--d15S100} and {\bf AT20G--d15S50}, if the flux limit is
100\,mJy and 50\,mJy respectively. For these sources, we were able to
determine the spectral index at low frequencies by exploiting ATCA
5--GHz measurements and low--frequency surveys in the southern sky:
the NVSS at 1.4\,GHz for $\delta>-40^{\circ}$; the Sydney University
Molonglo Sky Survey \citep[SUMSS][]{mau03} and the Molonglo Galactic
Plane Survey \citep[MGPS][]{mur07} at 843\,MHz for
$\delta\le-40^{\circ}$. Whereas 5--GHz ATCA measurements were not
available, we searched for the source counterparts in the Southern
Parkes-MIT-NRAO (PMN) survey \citep{wri94,wri96}. In AT20G--d15S100
we obtained the spectral index for 2158 sources, more than 98\% of
the sub--sample: there were 395 sources with steep spectrum and 1763
with flat spectrum ($\sim82\%$), of which 471 with inverted spectrum
$\alpha\ge0.3$. In Table\,\ref{t0} we report the number of sources
with 5--GHz measurements and the number of steep-- and
flat--spectrum sources in the two sub--samples of the AT20G. The
percentages of flat-- and steep--spectrum sources
agree with the ones given by \citet{mas10b} over a slightly
different and smaller area of AT20G (see their Figure 1), where they
found $\sim82\%$ and $\sim74\%$ of flat--spectrum sources with flux
density $\ge100$ and 50\,mJy respectively.

\begin{table}
\caption{Spectral information in the two almost--complete sub--samples
  of AT20G. Columns give the number of sources; number of sources with
  estimated 1--5\,GHz spectral index; number of flat-- and
  steep--spectrum sources.}
\centering
\begin{tabular}{cccccc}
\hline
Sample & $N_{tot}$ & $N(\alpha_1^5)$ & flat & steep \\
\hline
AT20G--d15S100 & 2195 & 2158 & 1763 & 395 \\
AT20G--d15S50  & 3807 & 3699 & 2821 & 878 \\
\hline
\end{tabular}
\label{t0}
\end{table}

\item At $\nu\simeq30$\,GHz various CMB experiments have provided
  samples of extragalactic radio sources, giving estimates of the
  number counts at different ranges of flux densities: the Cosmic
  Background Imager \citep[CBI;][]{mas03} in the range 5--50\,mJy; the
  Degree Angular Scale Interferometer \citep[DASI;][]{kov02} for
  $S\ga100\,$mJy; the Very Small Array \citep[VSA;][]{cle05} in the
  range 20--114\,mJy; the Sunyaev--Zel'dovich Array
    \citep[SZA;][]{muc10} in the range 0.7--15\,mJy.

\item The Wilkinson Microwave Anisotropy Probe (WMAP) carried out
  all-sky surveys at 23, 33, 41, 61, and 94\,GHz and provided a
  catalogue of ERS at a completeness levels of $\ga1\,$Jy. Analyses
  from the WMAP team have yielded 390 point sources in the five--year
  data \citep{wri09}, whereas 62 new point sources are found in the
  seven--year data \citep{gol10}. WMAP five--year maps has been also
  analyzed by \citet{mas09} that detected 516 point sources, 457 of
  which were previously identified as extragalactic sources.

\item Data on ERS are also present for frequencies $\nu\ga100$\,GHz:
  the South Pole Telescope \citep[SPT;][]{vie09} carried out a survey
  of ERS at 1.4 and 2.0\,mm wavelengths with arcmin resolution and mJy
  depth over an area of 87\,deg$^2$; the Atacama Cosmology Telescope
  \citep[ACT;][]{mar10} provided a catalog of 157 sources with flux
  density between 15 and 1500\,mJy detected at 148\,GHz in an area of
  455\,deg$^2$.

\item The {\it Planck} ERCSC \citep{Planck11c} reported data on
  compact sources detected in the nine {\it Planck} frequency channels
  between 30 and 857 GHz during the first 1.6 full-sky surveys. The
  analysis of the {\it Planck} ERCSC data presented in
  \citet{Planck11i} is limited to a primary sample of 533 compact
  extragalactic sources at $\vert b\vert>5^{\circ}$, selected at 30
  GHz.  More than the $97\%$ of these compact objects have been
  identified in external, published catalogues of ERS at GHz
  frequencies (see the {\it Planck} ERCSC Explanatory Supplement, for
  more details). Moreover, this 30--GHz sample is found to be
  statistically complete down to a flux density of $\approx 0.9$-$1.0$
  Jy and 290 ERS are found at above this flux density limit. The {\it
  Planck Collaboration} has been able to measure the number counts
  at the {\it Planck} LFI (30, 44 and 70 GHz) and at the HFI
  frequencies of 100, 143 and 217 GHz, with an estimated completeness
  limits of 1.0, 1.5, 1.1, 0.9, 0.5, 0.4\,Jy respectively
  \citep{Planck11i}.

\end{itemize}


\section{Estimate of the break frequency in blazars}

Spectra for the synchrotron emission from a spherical and
homogeneous source have a peak due to the self--absorption of their
own radiation. The observed frequency at which the peak of the
self-absorption occurs depends on the magnetic field and the depth
of the source, and it can be computed by \citep{pac70}
\beq
\nu_{syn,\,abs}=C_{\alpha}\,\bigg({S_{syn,\,abs} \over {\rm
    Jy}}\bigg)^{2/5}\,\bigg({\theta \over {\rm
    mas}}\bigg)^{-4/5}\,\bigg({H \over {\rm
    mG}}\bigg)^{1/5}\,(1+z)^{1/5}\,\delta^{(1-2p/5)}\,,
\label{a1}
\eeq
where $\theta$ is the observed angular dimension of
the source and $\nu_{syn,\,abs}$ is measured in GHz.
The parameter $p$ depends on the emission
model and gives the enhancement of the observed flux due to the
beaming ($S_{obs}=\delta^{\,p-\alpha}\,S$). It is equal to 3 for a
moving isotropic source and 2 for a continuous jet (see
\citealt{ghi93} and \citealt{urr95} for detailed
discussion). In the analysis we used as reference value $p=3$.
We also assumed
that emitting electrons have a power--law energy distribution
$N(\gamma)=K\gamma^{-(1-2\alpha)}$, where $\alpha$ is the spectral
index of the optically thin synchrotron emission. The term
$C_{\alpha}$ in Eq.\,\ref{a1} depends on $\alpha$ by
\beq
C_{\alpha}=2c_1\Bigg[{4c_6(\alpha) \over
\tau_m(\alpha)c_5(\alpha)}\Bigg]^{2/5}\,,
\eeq
with $c_1=3e/4\pi m^3c^5$
and $\tau_m$ the optical depth of the source at $\nu_m$. The
functions $c_5(\alpha)$ and $c_6(\alpha)$ are provided in
\citet{pac70}.

As discussed in the text, the break frequency ($\nu_M$) in a
flat--spectrum source is approximately the synchrotron
self--absorption frequency $\nu_{syn,\,abs}$ for the innermost part of
the jet whose emission is observed at cm/mm wavelengths. If this
region is assumed to be homogeneous and spherical (with diameter $d$),
$\nu_M$ can be obtained from Eq.\,\ref{a1}. For a conical jet
geometry, the diameter is related to the distance from the AGN core
$r_M$ by $d=2r_M\tan(\phi/2)\simeq\phi r_M$, where $\phi$ is the
semiangle of the conical jet.

If the flux density for a flat--spectrum source is known at the
observational frequency $\nu_o$, the observed flux density at $\nu_M$
($S_M$) can be extrapolated from $\nu_o$ using a power law spectrum
(we are assuming that $S_M\simeq S_{syn,\,abs}$, i.e. the
contributions from other jet regions are negligible at the frequency
$\nu_M$):
\beq
S_M=C_S\,\nu_M^{\alpha_{fl}}~~~~~~~{\rm
  with}~~C_S=\nu_o^{-\alpha_{fl}}\,S(\nu_o)\,.
\label{a2}
\eeq

Moreover, we have seen from Eq.\,\ref{e3}--\ref{e4} that the magnetic
field in equipartition condition is
\beq
H_{eq}\simeq1.06\times10^{-11}\Bigg[{(L/{\rm Watt}) \over (V/{\rm
      Kpc}^3)}\Bigg]^{2/7} \,[{\rm mG}]\,.  
\eeq 
The total luminosity
of the source $L$ can be calculated by the integral of the observed
flux density: 
\beq 
L=\int_{\nu_{min}}^{\nu_{cut}}L(\nu)d\nu={4\pi
  D_L^2 \over (1+z)^{1+\alpha}}
\delta^{\alpha-p}\int_{\nu_{min}}^{\nu_{cut}}S(\nu)d\nu~~[{\rm Watt}]\,,
\eeq
where $D_L$ is the luminosity distance in Mpc, $\nu_{min}$ and
$\nu_{cut}$ give the frequency range where the source emission is
concentrated (we take $\nu_{min}=10$\,MHz and $\nu_{cut}=10^5$\,GHz,
in agreement with standard assumptions on synchtrotron emission in
blazar sources). We used the relation
between the luminosity emitted at a given frequency and the observed
flux density, which takes into account the K--correction and the
relativistic beaming effects on the flux. The integral on the flux
density is now expressed as a function of $S_M$ and $\nu_M$:
\begin{eqnarray}
\int_{\nu_{min}}^{\nu_{cut}}S(\nu)d\nu & = & \int_{\nu_{min}}^{\nu_{cut}}
S_M\bigg({\nu \over \nu_M}\bigg)^{\alpha}d\nu \nonumber \\
& = & {S_M\nu_M^{-\alpha} \over 1+\alpha}\nu_{cut}^{1+\alpha}
\bigg[1-\bigg({\nu_{min} \over \nu_{cut}}\bigg)^{1+\alpha}\bigg]
\end{eqnarray}
for $\alpha\ne-1$ [if $\alpha=-1$ the integral is equal to
$S_M\nu_M\ln(\nu_{cut}/\nu_{min})$. Below we assume this condition
is verified; it is easy to extend the calculation for
$\alpha=-1$]. Using Eq.\,\ref{a2} the total luminosity becomes
\beq
L=C_L\,{D_L^2\delta^{\alpha-p} \over (1+z)^{1+\alpha}}
\nu_M^{\alpha_{fl}-\alpha}~~[{\rm Watt}],
\eeq
where
\beq
C_L\simeq9.5\times10^{27}4\pi S(\nu_o)\nu_o^{-\alpha_{fl}}
{\nu_{cut}^{1+\alpha} \over 1+\alpha}\Bigg[1-\bigg({\nu_{min} \over
\nu_{cut}}\bigg)^{1+\alpha}\Bigg]\,.
\eeq
By assuming for simplicity that the source is spherical with diameter
$d\simeq0.1r_M$ \citep{kon81,ghi09}, we obtain the expression for
the magnetic field:
\beq
H=C_H\,{D_L^{4/7}\delta^{{2 \over 7}(\alpha-p)} \over
(1+z)^{{2 \over 7}(1+\alpha)}}
r_M^{-6/7}\nu_M^{2(\alpha_{fl}-\alpha)/7}~~[{\rm mG}],
\label{a3}
\eeq
where
\beq
C_H\simeq0.47\times10^{-6}C_L^{2/7}\,.
\eeq

The observed angular dimension of the source in Eq.\,\ref{a1} is
\beq
\theta=C_{\theta}(1+z^2)r_M/D_L~~[{\rm mas}]\,,
\label{a4}
\eeq
with $C_{\theta}\simeq2.1\times10^7$.

Finally, including Eq.\,\ref{a2}--\ref{a3}--\ref{a4} in Eq.\,\ref{a1},
the break frequency becomes
\beq
\nu_M=C(\alpha,\,\alpha_{\rm fl})\bigg[D_L^{\beta_D}
(1+z)^{\beta_z}
\delta^{\beta_{\delta}}r_M^{\beta_d}\bigg]^{1/\beta}\,,
\eeq
where
$C(\alpha,\,\alpha_{\rm fl})=(C_{\alpha}\,C_s^{2/5}C_{\theta}^{-4/5}C_H^{1/5})^{1/\beta}$,
and $\beta=1+{2 \over 35}\alpha-{16 \over 35}\alpha_{fl}$,
$\beta_D={32 \over 35}$,
$\beta_z={2\alpha-51 \over 35}$,
$\beta_d=-{34 \over 35}$,
$\beta_{\delta}=1+{2 \over 35}\alpha-{16 \over 35}p$.


\section{Physical quantities relevant for the estimate of the break
frequency in blazars spectra}
\label{ss3}

\subsection{Redshift distribution of blazars}

In \citet{mas10} the {\bf redshift distribution} of radio sources at
low frequencies is widely discussed, and we followed this paper to
derive the redshift distribution of flat--spectrum sources. Most of
the samples with redshift information do not distinguish between
steep-- and flat--spectrum sources. Spectral information are present
in the \citet{kuh81} catalogue, however: this sample comprises 518 ERS
to a 5--GHz flux density limit of 1\,Jy, over an area of 9.811 sr.
Based on the catalogued spectral indices, 299 sources are classified
as flat-spectrum; 212 of which are FSRQs (200 with measured redshift),
26 are BL\,Lacs (20 with measured redshift) and 61 are classified as
galaxies or with missing classification. Moreover, in the Parkes
quarter--Jy sample of flat--spectrum ERS \citep{jac02}, redshifts are
available for the 58\% of sources. From this sample, \citet{dez05}
have defined a complete sub--sample of 514 objects with flux limit of
0.25\,Jy, aiming at maximizing the fraction ($\sim75\%$) of ERS with
known redshift . This sub--sample includes 370 FSRQs (93\% with known
redshift) and 47 BL\,Lacs, of which only 21\% with known redshift.

\begin{figure}
\centering
\includegraphics[width=4cm]{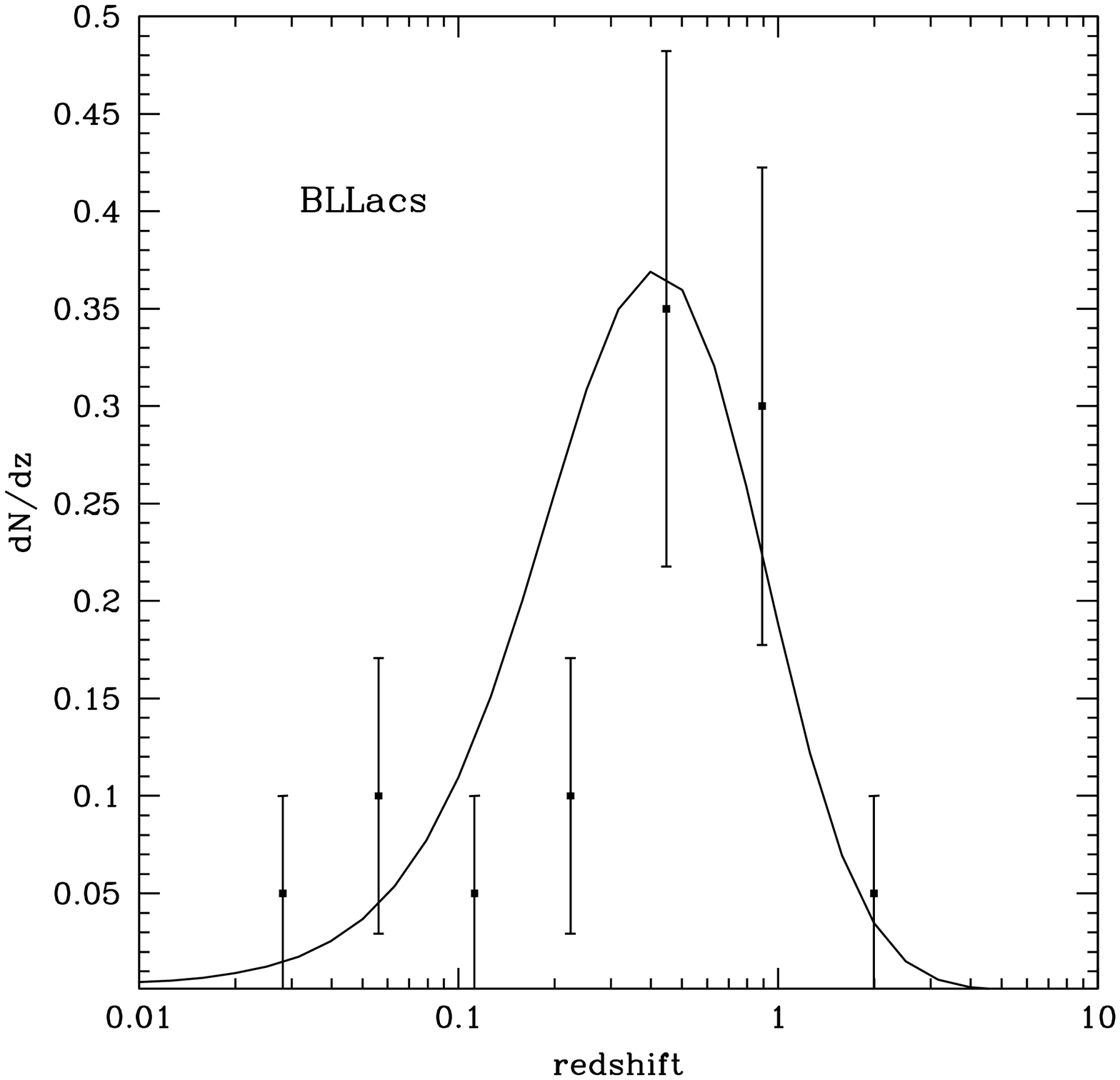}
\includegraphics[width=4cm]{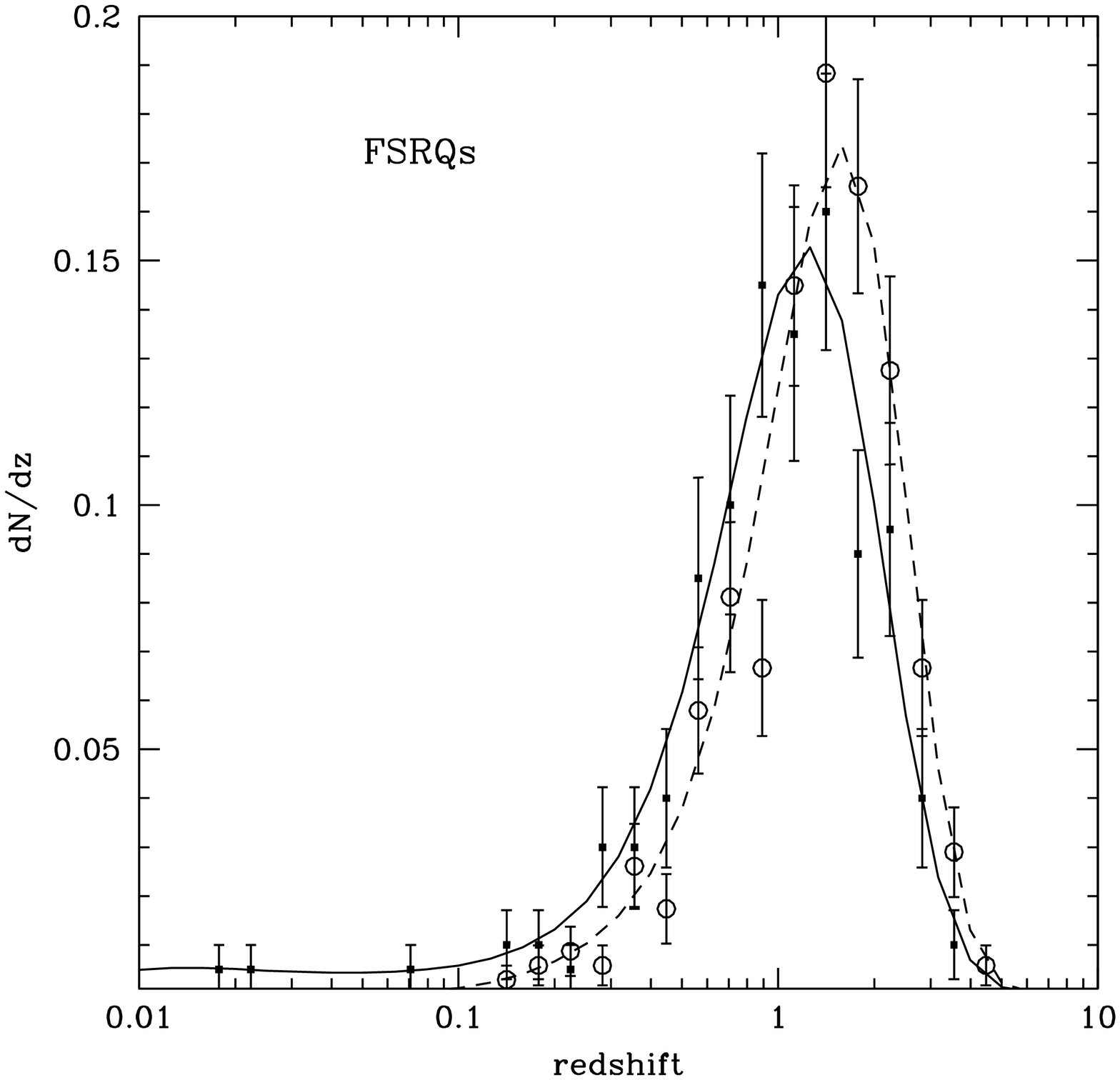}
\includegraphics[width=4cm]{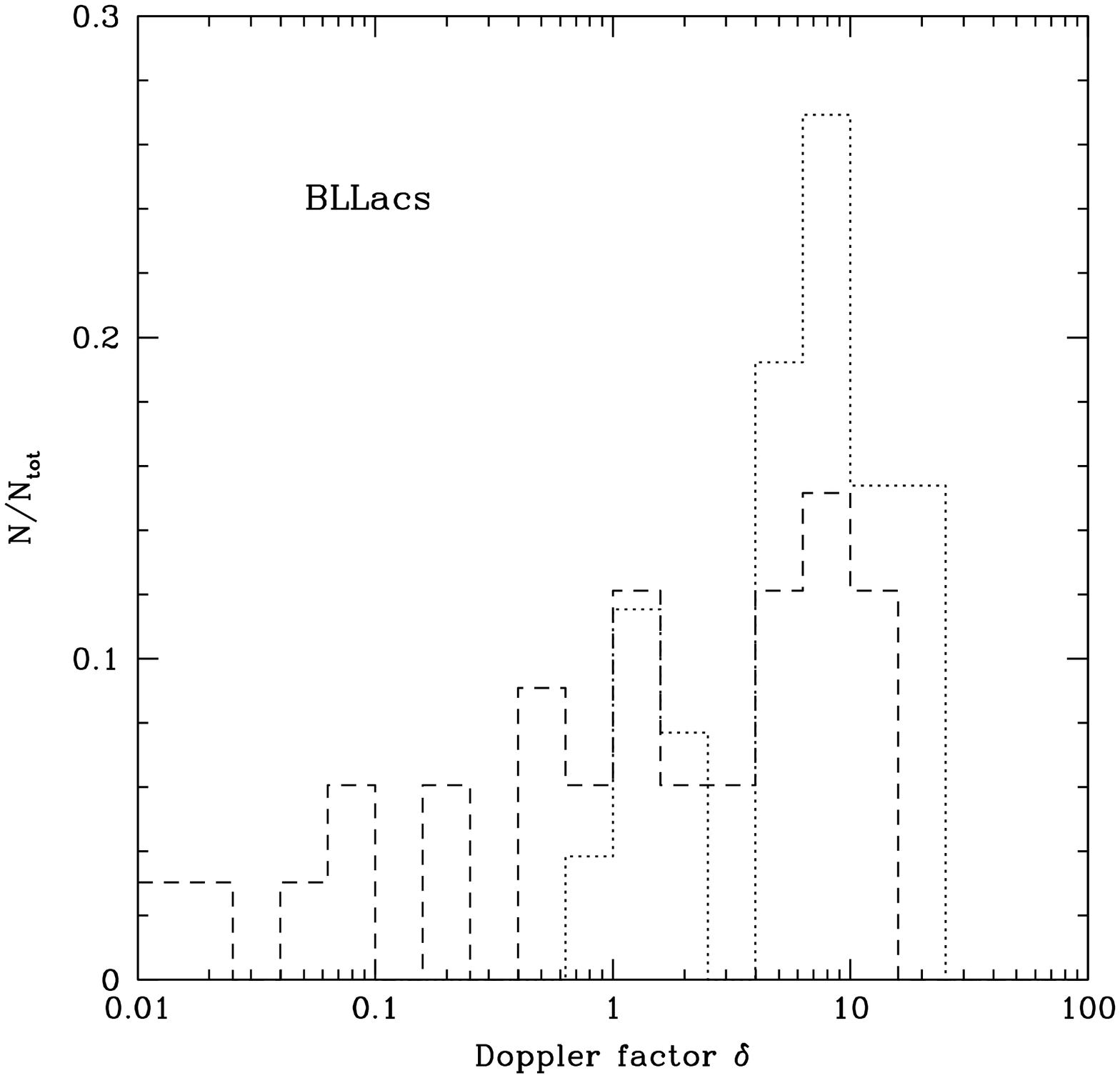}
\includegraphics[width=4cm]{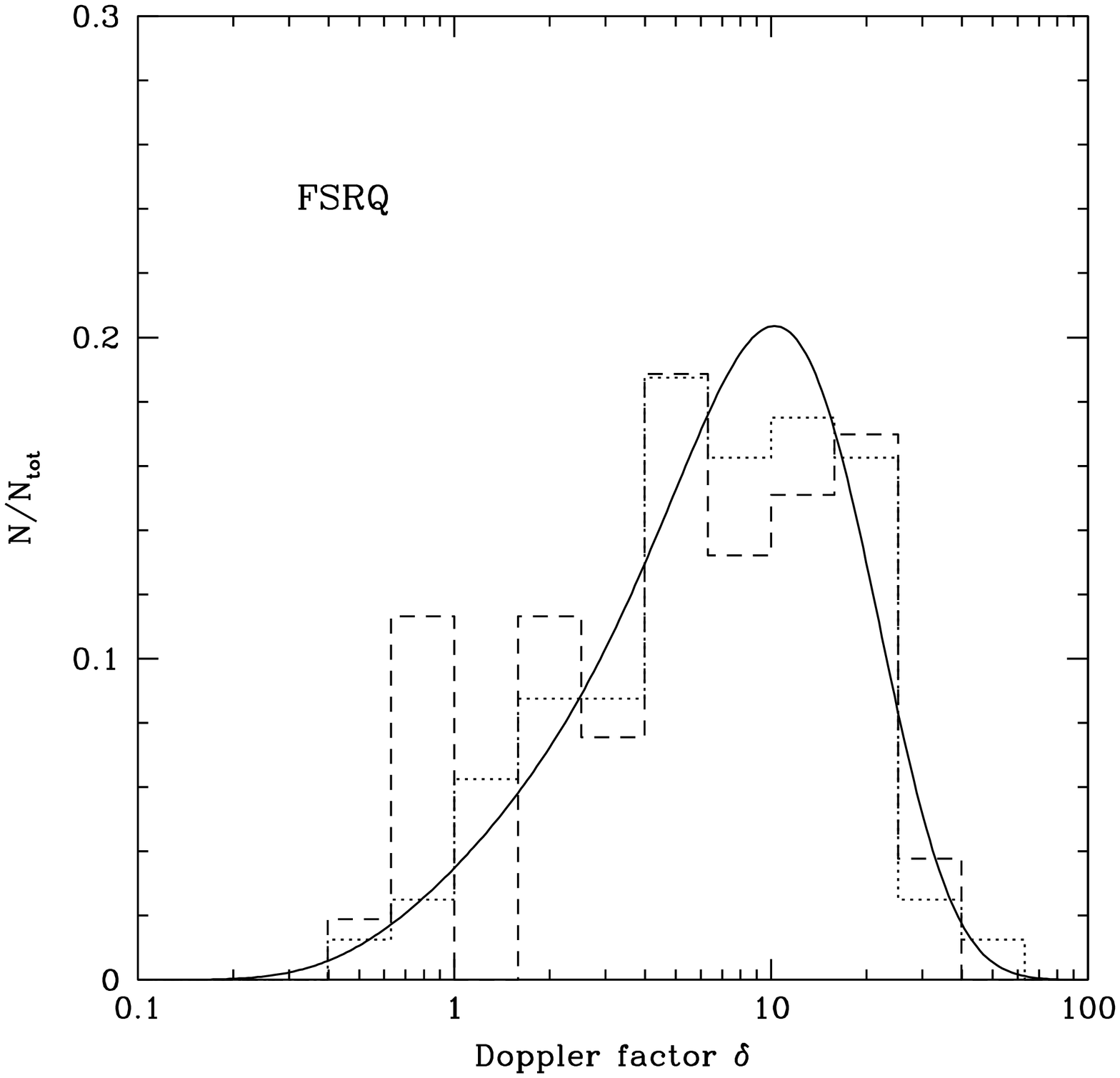}
\caption{{\it Upper panels}: redshift distributions of BL\,Lacs
({\it left panel}) and of FSRQs ({\it right panel}) from the samples
  by \citet{kuh81} (solid points) and \citet{jac02} (empty
  points), and the corresponding fits (solid and dashed lines, respectively).
  {\it Lower panels}: histograms of the Doppler factor as obtained by
  \citet{ghi93} (dashed lines) and \citet{gu09} (dotted lines) for
  BL\,Lacs ({\it left panel}) and core--dominated quasars ({\it
  right panel}). The solid line in the right panel is the fit of the two
  $\delta$ distributions for core--dominated quasars.}
\label{f6}
\end{figure}

In Fig.\,\ref{f6} we plot the redshift distributions of FSRQs and of
BL\,Lacs, and the fits we used for our predictions. For FSRQs only,
it has been possible to calculate them from both the samples: in
this case, the redshift distribution is observed to shift to higher
redshifts as the flux limit of the sample is lowered down, with the
peak of the distribution moving from $z\simeq1.2$ to $z\simeq 1.5$.
Moreover, Fig.\,\ref{f6} shows that the relative number of
low--redshift FSRQs is strongly reduced, if a fainter flux detection
limit is adopted. As for BL\,Lacs, the redshift distribution can be
obtained from the K$\ddot{u}$hr et al. catalogue, and for only 20
very bright objects. Because of the lack of information at faint
fluxes, this redshift distribution will be considered representative
also for BL Lac sources with flux density lower than the sample
limit. Owing to this, our predictions on number counts of ERS
discussed in Sect. 7 are, therefore, more uncertain when applied to
sources at $S< 0.1$ Jy.

\subsection{The Doppler factor in AGNs}

The estimate of the {\bf Doppler factor $\delta$} in AGNs is something
complex and model--dependent. In the framework of the synchrotron
self--Compton model \citep{mar87}, \citet{ghi93} calculated the
Doppler factor for a sample of 105 radio sources using VLBI
measurements of the core angular dimensions and radio fluxes. The
value of $\delta$ is calculated comparing the observed X--ray fluxes
with the ones predicted on the based of a homogeneous spherical
emitting model. In Fig.\,\ref{f6} we show the $\delta$ distribution
for the 53 core--dominated quasars and the 33 BL\,Lacs present in the
sample. These distributions can be compared with results from
\citet{gu09} where the Doppler factor is computed using the
K$\ddot{o}$nigl inhomogeneous jet model instead of the homogeneous
spherical model. Their sample consists of 128 sources, with 80
core--dominated quasars and 26 BL\,Lacs (37 quasars and 19 BL\,Lacs
are in common with the sample used by Ghisellini et al.).  The
$\delta$ distributions are similar in the case of core--dominated
quasars, with most of sources having $\delta$ between 1 and 30, as
expected for objects where the relativistic beamed emission is
dominant. For BL\,Lacs the results from \citet{ghi93} and \citet{gu09}
do not agree: the former find very low $\delta$ values,
extending from $10^{-2}$ to 10; in the latter the $\delta$
distribution is similar to the core--dominated quasars one. The
inhomogeneous model, in general, provides a better description of AGN
jet properties, but has the disavantage to involve more free
parameters than the homogeneous model. Note, however, that in the case
of the homogeneous model it is assumed that all the observed X--ray
flux is produced through inverse Compton scattering by the core
component dominanting at the radio frequency.  If part of the X--ray
flux is produced in other components or by some other mechanism, then
the computed $\delta$ is a lower limit.  For these reasons and for
greater simplicity, we assumed the same $\delta$ distribution for
BL\,Lacs and core--dominated quasars (in general for all the
flat--spectrum sources), described by the fit in Fig.\,\ref{f6}.


\end{document}